\documentclass[11pt]{article}
\pdfoutput=1
\usepackage{graphicx}
\usepackage{amsfonts}
\usepackage{amsmath}
\usepackage{amssymb}
\usepackage{bbold}
\usepackage{geometry}
\usepackage{longtable}
\usepackage{cite}

\begin{document}
\begin{titlepage}
\vspace*{-1cm}
\phantom{hep-ph/***}
\flushright
\hfil{RM3-TH/15-1}

\vskip 1.5cm
\begin{center}
\mathversion{bold}
{\LARGE\bf Lepton mixing from the interplay of the alternating group $A_5$ and CP
}\\[3mm]
\mathversion{normal}
\vskip .3cm
\end{center}
\vskip 0.5  cm
\begin{center}
{\large Andrea Di Iura}~$^{a)}$,
{\large Claudia Hagedorn}~$^{b)}$
{\large and Davide Meloni}~$^{a)}$
\\
\vskip .7cm
{\footnotesize
$^{a)}$~Dipartimento di Matematica e Fisica, Universit\`a  di Roma Tre;\\
INFN, Sezione di Roma Tre,\\
Via della Vasca Navale 84, 00146 Rome, Italy
\vskip .1cm
$^{b)}$~Excellence Cluster `Universe', Technische Universit\"{a}t M\"{u}nchen,\\ 
Boltzmannstra\ss e 2, 85748 Garching, Germany
\vskip .5cm
\begin{minipage}[l]{.9\textwidth}
\begin{center} 
\textit{E-mail:} 
\tt{diiura@fis.uniroma3.it}, \tt{claudia.hagedorn@ph.tum.de}, \tt{meloni@fis.uniroma3.it}
\end{center}
\end{minipage}
}
\end{center}
\vskip 1cm
\begin{abstract}

Assuming three generations of Majorana neutrinos, we study the different mixing patterns that arise from the non-trivial breaking of the flavor group $A_5$ and CP
to the residual symmetries $Z_3$, $Z_5$ or $Z_2 \times Z_2$ in the charged lepton and to $Z_2 \times CP$ in the neutrino sector.
All patterns contain only one free parameter $\theta$ and thus mixing angles as well as the Dirac and the two Majorana phases 
are strongly correlated. We perform an analytical and a numerical study of all possible 
mixing patterns.
It turns out that only four patterns can describe the experimentally measured values of the mixing angles for a
particular choice of $\theta$ well. All of them predict trivial Majorana phases, while the Dirac phase $\delta$ is maximal for two patterns and trivial
for the two remaining ones. If $\delta$ is maximal, also the atmospheric mixing angle is fixed to be maximal.

\end{abstract}
\end{titlepage}
\setcounter{footnote}{0}

\section{Introduction}
\label{intro}

By now all three lepton mixing angles have been measured with a certain degree of precision \cite{nufit} 
(see also \cite{otherglobal_1, otherglobal_2})
\begin{equation}
\!\!\!\!\!\!\!\sin^2 \theta_{13}=0.0218 (9)  \, ^{+ 0.0010 (1)} _{-0.0010} \;\; , \;\; \sin^2 \theta_{12} = 0.304 \, ^{+ 0.013} _{-0.012} \;\; , \;\;  \sin^2 \theta_{23} = 
\left\{ \begin{array}{c}
0.452 \, ^{+ 0.052} _{-0.028} \\[0.05in]
\left(   0.579 \, ^{+ 0.025} _{-0.037}  \right) 
\end{array}
\right.
\label{nufitresults}
\end{equation}
for a normal ordering (NO) and in brackets for an inverted ordering (IO) of the neutrino masses, respectively. 
Assuming that neutrinos are Majorana particles and none of them to be massless, the Pontecorvo-Maki-Nakagawa-Sakata (PMNS)
mixing matrix $U_{PMNS}$ that encodes lepton mixing does not only contain the three mixing angles $\theta_{12}$, $\theta_{13}$ and $\theta_{23}$,
but also three phases: the Dirac phase $\delta$ and the two Majorana phases $\alpha$ and 
$\beta$. The former can be measured in neutrino oscillation experiments, while a linear combination of the latter can be 
 accessible in neutrinoless double beta decay experiments (for a review on leptonic CP violation see \cite{Brancoreview}). No direct signals of CP 
 violation have been observed in the lepton sector and recent global fits only show a weak indication (below the $3 \, \sigma$ significance) of a non-trivial value of the Dirac phase $\delta$ \cite{nufit}.

Many approaches have been pursued in order to describe the data on lepton mixing. A particularly promising Ansatz assumes the existence
of a flavor symmetry $G_f$, usually finite, non-abelian and discrete, that is broken to different residual groups $G_e$ and $G_\nu$ in the charged lepton 
and neutrino sectors, respectively (for reviews see \cite{reviews,review_math}).
In (the most predictive version of) such a framework all three mixing angles together with the Dirac phase $\delta$
can be fixed by the symmetries of the theory $G_f$, $G_e$ and $G_\nu$, if the three generations of left-handed (LH) leptons are assigned to an irreducible three-dimensional representation
{\bf 3} of the flavor group $G_f$, i.e. to a representation that cannot be decomposed further in $G_f$.\footnote{Since lepton mixing like quark mixing only regards LH fields, we do not need
to specify in this approach how right-handed (RH) charged leptons, and possibly RH neutrinos, transform under $G_f$, unless we construct an explicit model in which this approach is
realized.} We note that this approach does not constrain lepton masses and 
thus all statements made on the predictive power regarding lepton mixing angles and the Dirac phase $\delta$ are valid up to possible 
permutations of rows and columns of the PMNS mixing matrix. For neutrinos being Majorana particles
surveys of finite, non-abelian and discrete subgroups of $SU(3)$ and $U(3)$, see \cite{Grimus2014} and references therein, have shown that symmetries
giving rise to mixing angles which are in good agreement with experimental data in general lead to trimaximal (TM) mixing \cite{TM} (and thus $\sin^2 \theta_{12} \gtrsim 1/3$)
and a trivial Dirac phase $\delta=0, \pi$. 

An extension of this approach that involves also CP as symmetry has been proposed in \cite{GfCPgeneral} (see also \cite{GfCPHD,ChenCP} as well as \cite{GrimusRebelo}).
In this case, also the CP symmetry acts in general in a non-trivial way on the flavor space \cite{Ecker} and conditions have to be fulfilled in order for the theory to be consistent.
The residual groups $G_e$ and $G_\nu$ are chosen as follows: $G_e$ is an abelian subgroup of $G_f$ with three or more elements and $G_\nu$ is the direct product
of a $Z_2$ group contained in $G_f$ and the CP symmetry. Thus, the form of these groups 
is similar to the one in the approach without CP. 
The advantages of the extension with CP are threefold: Majorana phases are also predicted, the Dirac phase does not need to be trivial, if the lepton mixing angles are accommodated well, and 
the mixing pattern contains one free parameter $\theta$. The latter allows for a richer structure of patterns that are in good agreement with the experimental data. In particular, mixing does not
need to be TM. At the same time, being governed by only one free parameter all mixing parameters are strongly correlated. Such correlations 
can be testable and/or distinguishable at future facilities \cite{sumrules_future}. The value(s) of this free parameter that admit(s) a reasonable agreement with the
experimental data is (are) not fixed by the approach itself, but has (have) to be achieved in a concrete model (see \cite{GfCPmodels} for several successful models with different symmetries $G_f$ and CP).
This approach has already been studied for a variety of flavor symmetries: $A_4$ and $S_4$ \cite{GfCPgeneral,GfCPmodels}, $\Delta (48)$ \cite{D48CP}, $\Delta (96)$ \cite{D96CP}
as well as $\Delta (3 \, n^2)$ and $\Delta (6 \, n^2)$ with general $n$ \cite{D3n2D6n2CP}.\footnote{Note that a variant of this approach has been considered for $G_f=\Delta (6 \, n^2)$
in which the residual symmetry in the neutrino sector is a Klein group contained in the flavor group and a CP symmetry \cite{D6n2CP_var}.}

Here we would like to consider $A_5$ as flavor group. This group has already been employed as flavor symmetry \cite{A5papers,A5FeruglioParis,modular,A5papers2}. In particular, it has been shown 
to give rise to the so-called ``golden ratio" (GR) mixing pattern, $\sin^2 \theta_{23}=1/2$, $\theta_{13}=0$ and $\tan\theta_{12}=1/\phi$
with $\phi= (1+\sqrt{5})/2 \approx 1.618$ so that $\sin^2 \theta_{12} \approx 0.276$.\footnote{Different 
versions of the GR mixing pattern are known in the literature that lead
to different predictions for the solar mixing angle in terms of the golden ratio \cite{modular,GRother, GRotherLam}. These are based on different flavor symmetries.}
Very recently, predictions of CP phases have been discussed in a scenario with $A_5$ as flavor group and a CP symmetry \cite{Everett_GfCP}. 
Since the authors assume a Klein group
and a CP symmetry to be preserved in the neutrino sector, the mixing angles are fixed to the values of the GR mixing pattern, while possible values of the two Majorana phases
depend on the CP transformation that is preserved.
The latter is not constrained to correspond to an automorphism of the flavor group $A_5$ and thus results obtained in \cite{Everett_GfCP} differ from ours. In \cite{twoCPs}
an Ansatz has been pursued in which two CP symmetries are present as residual symmetries in the neutrino sector. The combination of these two
leads in general to a symmetry acting on the flavor space only. Under certain conditions this can be a transformation belonging to a
finite, non-abelian and discrete group. Therefore, if the latter is the alternating group $A_5$ results of our study can also
be achieved using the Ansatz with two CP transformations in the neutrino sector.

In the present paper we analyze the scenario with the flavor symmetry $A_5$ and a CP symmetry comprehensively, since we consider all CP symmetries 
that correspond to involutive `class-inverting' automorphisms of $A_5$, all possibilities for $G_e$, i.e. $G_e=Z_3$, $G_e=Z_5$ as well as $G_e=Z_2 \times Z_2$,
and all possible $Z_2$ subgroups of $A_5$ as residual flavor symmetry in the neutrino sector. All these combinations of symmetries together
with all possible permutations of the rows and columns of the PMNS mixing matrix are subject to an analytical and a numerical study. In particular,
we perform a $\chi^2$ analysis using the results of the mixing angles from the global fit \cite{nufit}.
As outcome we only find four patterns that admit a reasonable agreement with the experimental
data, i.e. at the $3 \, \sigma$ level or better, for a particular choice of the parameter $\theta$. Two of these four patterns predict
a maximal Dirac phase together with maximal atmospheric mixing, while the other two ones lead to a trivial Dirac phase and in general
non-maximal $\theta_{23}$. Majorana phases are trivial for all four patterns. Thus, two out of these four lead to no CP violation.
This fact can be traced back to the existence of an accidental CP symmetry, common to the charged lepton and neutrino sectors, as we discuss. 
As regards the reactor and the solar mixing angles, we note that $\theta_{13}$ is in general accommodated well, whereas $\theta_{12}$
is subject to non-trivial constraints in all four cases: two of the four patterns give rise to a lower bound $\sin^2 \theta_{12} \gtrsim 0.276$, the value
of the GR mixing pattern, one incorporates TM mixing and thus $\sin^2\theta_{12} \gtrsim 1/3$ and the remaining one entails 
an upper limit $\sin^2\theta_{12} \lesssim 1- \phi^2/4 \approx 0.345$, a value that is associated with a different version of the GR mixing pattern \cite{GRother}.

The paper is organized as follows: in section \ref{sec2} we recapitulate the approach with a flavor and a CP symmetry as well as
the main features of the group $A_5$. We also discuss the admitted CP transformations and relegate further details regarding their
relation to the automorphisms of $A_5$ and their nature to appendix \ref{appA}. Section \ref{sec3} contains the analytical study of all patterns that can lead to a good agreement 
with the experimental data as well as the results of our $\chi^2$ analysis. We summarize our main results in section \ref{summ}.
Besides appendix \ref{appA} we include appendix \ref{appB}
that contains our definitions of mixing angles, CP phases and corresponding CP invariants $J_{CP}$, $I_1$ and $I_2$.

\mathversion{bold}
\section{Approach}
\mathversion{normal}
\label{sec2}

We briefly recapitulate the essential ingredients of the approach \cite{GfCPgeneral} and summarize the necessary information on the group $A_5$.
We list the candidates of generators of residual flavor symmetries in the charged lepton and neutrino sectors as well as the CP transformations.
At the end of this section we also comment on the possible presence of an accidental CP symmetry common to charged leptons and neutrinos. 
Since we focus on the case in which the generations of LH leptons are assigned to an irreducible three-dimensional representation ${\bf 3}$
of the flavor group, the CP transformation $X$ as well as the elements of the flavor symmetry are represented by  
(unitary, complex) three-by-three matrices in the following. 

Let us consider a theory with a flavor group $G_f=A_5$ combined with a CP symmetry that in general also acts
non-trivially on the flavor space \cite{Ecker,GrimusRebelo}. The CP transformation $X$ is a unitary and
symmetric matrix
\begin{equation}
\label{Xcon}
X X^\dagger = X X^\star = 1 \; .
\end{equation}
The requirement that $X$, the CP transformation associated with the CP symmetry preserved in the neutrino sector and subject to the condition in (\ref{XZcon}), must be a symmetric matrix
has been shown in \cite{GfCPgeneral} to be necessary, since otherwise the neutrino mass spectrum would be partially degenerate and, consequently, inconsistent with
experimental observations \cite{nufit}. In order to ensure a consistent combination of the flavor and CP symmetry we require that the subsequent action of the CP transformation, 
an element of the flavor group and the CP transformation is equivalent to the action of an (in general different) element of the flavor 
group
\begin{equation}
\label{XGfcon}
(X^{-1} A X)^\star = A^\prime\,,
\end{equation}
with $A$ and $A^\prime$ representing (different) elements of $G_f$.\footnote{We use throughout this paper lowercase letters 
for the abstract elements of the flavor group $A_5$ and capital letters for the 
matrix representatives (in the representation ${\bf 3}$).} As shown in \cite{GrimusRebelo,GfCPHD,ChenCP}, a CP transformation
corresponds to an automorphism of the flavor group. In particular, our request that $X$ fulfills (\ref{Xcon}) renders this automorphism involutive.
Following the discussion in \cite{ChenCP} this automorphism should be class-inverting, i.e. the image of the element $g$ under the automorphism
has to lie in the same class as the inverse of $g$. This is guaranteed for the three-dimensional representation ${\bf 3}$ by the fulfillment of the condition in (\ref{XGfcon}). As
we show in appendix \ref{appA} the automorphisms corresponding to the CP transformations we consider in our analysis are also class-inverting when
acting on the other representations of the flavor group $A_5$.  

 The residual symmetry in the neutrino sector is assumed to be
the direct product of a $Z_2$ symmetry contained in the flavor group and the CP symmetry. Thus,  the matrix $Z$ representing the generator of the former symmetry
and the CP transformation $X$ have to fulfill
\begin{equation}
\label{XZcon}
X Z^\star - Z X = 0 \,,
\end{equation}
which is a particular case of the condition in (\ref{XGfcon}). We note that the presence of the residual symmetries given by $Z$ and $X$ implies the existence of a second 
CP transformation $\tilde{Y}=Z X$ in the neutrino sector that fulfills the same conditions in (\ref{Xcon})-(\ref{XZcon}) as the CP transformation $X$ \cite{GfCPgeneral}.

In the charged lepton sector, in contrast, we take as residual group an abelian
subgroup of the flavor symmetry that offers the possibility to distinguish among the three generations of charged leptons, i.e. this group has to have
at least three different elements. 
The residual group
can be described with a set of generators $Q_i$, $i=1,2,...$ that commute. 

The derivation of lepton mixing in this scenario has already been presented
in detail in \cite{GfCPgeneral} and we only mention it briefly. In the charged lepton sector, the residual group generated by $Q_i$ constrains the charged lepton
mass matrix $m_l$, here given in the right-left basis,
\begin{equation}
Q_i^\dagger \, m_l^\dagger m_l \, Q_i = m_l^\dagger m_l \; .
\end{equation}
The matrices $Q_i$ are diagonalizable by the unitary matrix $U_e$, i.e. $U_e^\dagger Q_i U_e$ is diagonal, which is determined up to the ordering of its 
columns and possible overall phases of the single columns. As a consequence, also the combination
\begin{equation}
U^\dagger_e \, m_l^\dagger m_l \, U_e \;\;\ \mbox{is diagonal} \; .
\end{equation}
Thus, $U_e$ diagonalizes the charged lepton mass matrix as regards LH charged leptons. In the neutrino sector, the light neutrino mass matrix $m_\nu$ is
subject to the following conditions\footnote{We do not need to specify the generation mechanism of neutrino masses unless we construct an explicit model.
So, this mass matrix can arise from integrating out heavy RH neutrinos, 
from Higgs $SU(2)_L$ triplets acquiring a vacuum expectation value, etc..}
\begin{equation}
Z^T \, m_\nu \, Z = m_\nu \;\;\; \mbox{and} \;\;\; X \, m_\nu \, X = m_\nu^\star \; ,
\end{equation}
if the $Z_2$ and CP symmetry are imposed. Without loss of generality we can choose a basis such that
\begin{equation}
\label{Omegacon}
X=\Omega \, \Omega^T \;\;\; \mbox{and} \;\;\; \Omega^\dagger \, Z \, \Omega = \mbox{diag} 
\left( (-1)^{z_1}, (-1)^{z_2}, (-1)^{z_3} \right)\,,
\end{equation}
with $\Omega$ being unitary and $z_i=0,1$. Since $Z$ generates a $Z_2$ symmetry, two of the three parameters $z_i$ have to coincide. 
The matrix combination $\Omega^T \, m_\nu \, \Omega$ is then real and block-diagonal
and thus can be diagonalized by a rotation $R_{ij} (\theta)$ in the $(ij)$-plane through an angle $\theta$ that is determined by the
matrix entries of $\Omega^T \, m_\nu \, \Omega$. Their actual values are in general not predicted in this approach and thus
$\theta$ is taken to be a free parameter in the interval between $0$ and $\pi$ in the following. The $(ij)$-plane is fixed by the degenerate sub-sector
of $\Omega^\dagger \, Z \, \Omega$, i.e. the two $z_i$ and $z_j$ that are equal. The positiveness of the light neutrino masses (a vanishing 
neutrino mass can also be included) is ensured by the diagonal matrix $K_\nu$ with entries $\pm 1$ and $\pm i$ on its diagonal. 
So, the contribution to lepton mixing from the neutrino sector is given by
\begin{equation}
\label{formUnu}
U_\nu= \Omega \, R_{ij} (\theta) \, K_\nu \,,
\end{equation}
and the PMNS mixing matrix resulting from this approach reads
\begin{equation}
\label{formUPMNS}
U_{PMNS}=U^\dagger_e \, \Omega \, R_{ij} (\theta) \, K_\nu \; .
\end{equation}
It is important to note that this mixing matrix is only determined up to permutations of its rows and columns (and unphysical phases), since
this approach does not make any predictions concerning the mass spectrum of charged leptons and neutrinos. For example, if not embedded 
in a model context, see e.g. first reference in \cite{GfCPmodels}, one cannot predict whether neutrinos follow NO or IO. 
It is also worth to emphasize that a mixing matrix of the form in (\ref{formUPMNS}) has one column that is determined by group theory only
and that does not depend on the free parameter $\theta$.

We note also that two tuples $(Q,Z,X)$ and $(Q^\prime,Z^\prime,X^\prime)$ lead to the same physical results, if the generators of the symmetries
are related by a similarity transformation $\tilde{\Omega}$
\begin{equation}
\tilde{\Omega}^\dagger \, Q^\prime \, \tilde{\Omega} = Q \;\;\; , \;\;\; \tilde{\Omega}^\dagger \, Z^\prime \, \tilde{\Omega} = Z \;\;\; \mbox{and} \;\;\; \tilde{\Omega}^\dagger \, X^\prime \, \tilde{\Omega}^\star = X \; .
\end{equation}

The alternating group $A_5$ describes the even permutations of five distinct objects. It is isomorphic to the icosahedral rotation group $I$.
It has 60 different elements, organized in five conjugacy classes, and, thus, possesses five irreducible representations: ${\bf 1}$, ${\bf 3}$, ${\bf 3'}$, ${\bf 4}$ and ${\bf 5}$.
All representations apart from the singlet are faithful.\footnote{A representation is called faithful, if all elements of the group 
are represented by different matrices in this representation. So, in the case of $A_5$ the 60 group elements are represented by 60 different matrices in the representations ${\bf 3}$, ${\bf 3'}$,
${\bf 4}$ and ${\bf 5}$.} The group $A_5$ can be generated with two generators $s$ and $t$
that fulfill the relations
\begin{equation}
s^2=e \;\; , \;\; t^5=e \;\;\; \mbox{and} \;\;\; (s t)^3=e\,,
\end{equation}
with $e$ denoting the neutral element of $A_5$.
Since we would like to assign LH leptons to triplets, we are particularly interested in the representations ${\bf 3}$ and ${\bf 3'}$.
The explicit form of the generators $s$ and $t$ in the representation ${\bf 3}$ can be chosen as \cite{A5FeruglioParis}
\begin{equation}
\label{ST3}
S=\frac{1}{\sqrt{5}}
\left(
\begin{array}{ccc}
1&\sqrt{2}&\sqrt{2}\\
\sqrt{2}&-\phi&1/\phi\\
\sqrt{2}&1/\phi&-\phi\\
\end{array}
\right) \;\;\; \mbox{and} \;\;\;
T=\left(
\begin{array}{ccc}
1&0&0\\
0&e^{i \, \Phi}&0\\
0&0&e^{4 \, i \, \Phi}
\end{array}
\right) \,,
\end{equation}
with 
\begin{equation}
\phi=\frac 12 \, (1+\sqrt{5}) \;\;\; \mbox{and} \;\;\; \Phi= \frac{2 \pi}{5} \; .
\end{equation}
The generators in ${\bf 3'}$ are easily obtained from $S$ and $T$ in (\ref{ST3}) by using the combination $T^2\, S\, T^3\, S\, T^2$
and $T^2$ as generators, see also \cite{modular}. This shows immediately that the set of all matrices describing the representations ${\bf 3}$
and ${\bf 3'}$ is the same and thus all conclusions obtained in a comprehensive study of mixing using 
the representation ${\bf 3}$ also hold for ${\bf 3'}$. Consequently,
it is irrelevant for our analysis whether LH leptons are in ${\bf 3}$ or ${\bf 3'}$ of $A_5$ and, without loss of generality, we
assume in the following that LH leptons transform as ${\bf 3}$ of $A_5$.

The group $A_5$ has several subgroups. In particular, the group contains 15 elements that generate a $Z_2$ symmetry 
which give rise to five distinct Klein groups. These are, like the $Z_2$ generating elements themselves, all conjugate to each other.
Since we make explicit use of these elements we mention them here\\[0.01in]

\begin{tabular}{lllll}
$v_1=s \; ,$&$v_2=s t^2 s t^3 s t^2 \; ,$&$v_3=t^2 s t^3 s t^2 \; ,$&$v_4= t^4 s t \; ,$&$v_5=s t^3 s t^2 s \; ,$
\\
$v_6=t^2 s t^3 s t s \; ,$&$v_7=t s t^4 \; ,$&$v_8=s t^2 s t^3 s \; ,$&$v_9= s t s t^3 s t^2 \; ,$&$v_{10}=s t^2 s t \; ,$
\\
$v_{11}=t^2 s t^3 \; ,$&$v_{12}=t s t^3 s t^2 s \; ,$&$v_{13}=t s t^2 s \; ,$&$v_{14}=t^3 s t^2 \; ,$&$v_{15}=s t^2 s t^3 s t \; .$
\\[0.15in]
\end{tabular}
\noindent They form the Klein groups $K_i$
\begin{eqnarray}\nonumber
&&K_1= \{ v_1, v_2, v_3, e \} \;\; , \;\; K_2= \{ v_4, v_5, v_6, e \} 
\\ \label{Kgens}
&&K_3= \{ v_7, v_8, v_9, e \} \;\; , \;\; K_4= \{ v_{10}, v_{11}, v_{12}, e \} \;\;\; \mbox{and} \;\;\;  K_5= \{ v_{13}, v_{14}, v_{15}, e \} \; ,
\end{eqnarray}
see also \cite{modular}.
Furthermore, there are ten $Z_3$ and six $Z_5$ subgroups.  Also these are all conjugate to each other. 
For the complete list of generating elements of these subgroups see appendix B in \cite{modular}.
So, three different types of groups, $Z_3$, $Z_5$ and $Z_2 \times Z_2$, 
can function as residual symmetry $G_e$ in the charged lepton sector.

The form of the CP transformations we consider is
\begin{equation}
\label{formX}
X= V \, X_0
\end{equation}
with $X_0$
\begin{equation}
X_0= \left(
\begin{array}{ccc}
1 & 0 & 0\\
0 & 0 & 1\\
0 & 1 & 0
\end{array}
\right) 
\end{equation}
and $V$ is the matrix representative of a $Z_2$ generating element,  i.e. $V^2=1$, or $V=1$. 
So, we find in total 16 different possible CP
transformations. All of them fulfill (\ref{Xcon}). Notice that the CP transformation $X_0$
gives rise to the so-called $\mu$-$\tau$ reflection symmetry \cite{mutaureflection}
whose phenomenological consequences have been studied in detail in the literature.

As one can check, it holds for $X_0$
\begin{equation}
\label{X0ST}
(X_0^{-1} S X_0)^\star = S  \;\;\; \mbox{and} \;\;\; (X_0^{-1} T X_0)^\star = T  \; .
\end{equation}
Thus, for the generators $S$ and $T$ in the representation ${\bf 3}$ the condition in (\ref{XGfcon}) is valid for $A=A^\prime$.
As a consequence, the corresponding automorphism is the trivial one
\begin{equation}
\label{autoX0}
s \;\;\; \rightarrow \;\;\; s \;\;\;\;\; \mbox{and} \;\;\;\;\; t \;\;\; \rightarrow \;\;\; t \; .
\end{equation} 
The CP transformations $X=V \, X_0$ correspond to different inner automorphisms of the group $A_5$  (i.e. automorphisms whose action on the elements of the group
can be represented by a similarity transformation with a(nother) group element) that map the generators $s$ and $t$ in the following way 
\begin{equation}
\label{autoX}
s \;\;\; \rightarrow \;\;\; v \, s \, v^{-1} \;\;\;\;\; \mbox{and} \;\;\;\;\; t \;\;\; \rightarrow \;\;\; v \, t \, v^{-1} \; .
\end{equation}
The automorphism group of $A_5$ is the symmetric group $S_5$ and the group of inner automorphisms is isomorphic to $A_5$ itself. 
As we show in appendix \ref{appA}
the 16 different CP transformations we consider correspond to the 16 class-inverting involutive automorphisms of $A_5$. 
We thus discuss all CP transformations that can be consistently imposed according to \cite{ChenCP} and that fulfill the requirement in (\ref{Xcon}).

A last condition that needs to be examined is the constraint in (\ref{XZcon}), namely whether the $Z_2$ generator 
commutes with the chosen CP transformation. We find that for each $Z$ four different possible CP transformations $X$ are admitted: for $Z$
being one of the non-trivial elements of the Klein group $K_i$, $i=1, ..., 5$, the CP transformation $X= V \, X_0$ with $V$ belonging to the same
Klein group $K_i$ (this time $V=1$ is included) is a viable choice. 
Taking into account that the $Z_2$ generator $Z$ and the CP transformation
$X$ automatically imply the existence of a further CP transformation $\tilde{Y}$, $\tilde{Y}=Z X$,
we can reduce the number of independent choices of CP transformations for each $Z_2$ generator $Z$ to two. 

Eventually, we mention that it can happen that an accidental CP symmetry $Y$, common to the charged lepton and the neutrino sectors, exists
that leads to trivial CP phases. This can be checked by searching for a transformation $Y$ that fulfills the 
following constraints
\begin{equation}
\label{CPacc1}
Y^\star \, m^\dagger_l m_l \, Y = (m^\dagger_l m_l)^\star \;\;\; \mbox{and} \;\;\; Y \, m_\nu \, Y = m_\nu^\star \; .
\end{equation}
These are equivalent to the conditions involving the generators of the different symmetries
\begin{eqnarray}
&&Q_i \, Y - Y \, Q_i^T=0 \; ,
\\ \label{CPacc2}
&&Z Y - Y Z^\star=0 \;\;\; \mbox{and} \;\;\; X Y^\star -Y X^\star =0\,,
\end{eqnarray}
with $Y$ being a diagonal and real matrix in the neutrino mass basis, i.e. $U_\nu^\dagger \, Y \, U_\nu^\star$ is diagonal and real.
For details see \cite{GfCPgeneral}.

\mathversion{bold}
\section{Lepton mixing}
\mathversion{normal}
\label{sec3}

In order to study mixing comprehensively, we analyze all possible combinations of residual symmetries in the 
charged lepton and neutrino sectors, i.e. all possible $G_e$ and $G_\nu$ ($Z_2$ generators and CP transformations $X$).
These can be expressed as tuples of generators $Q_i$ of $G_e$ and $Z$ and $X$ for $G_\nu$.
As mentioned in section \ref{sec2}, $G_e$ can be either a $Z_3$, $Z_5$ or a Klein group, while one of the 
15 different $Z_2$ symmetries that each can be 
consistently combined with four different CP transformations $X$ specify the residual group $G_\nu$. 
Instead of computing the mixing pattern for all these
we can greatly reduce the number of cases that we need to study by applying similarity transformations as well as the fact that
for a pair $Z$ and $X$ also the pair $Z$ and $\tilde{Y}=Z X$ leads to the same mixing pattern. We allow for all possible permutations of rows and columns
of the mixing matrix. Furthermore, we consider both possible neutrino mass orderings NO or IO in our (numerical) analysis.
We exclude all patterns that cannot accommodate the experimental
data on lepton mixing angles at the $3 \, \sigma$ level or better for (a) certain choice(s) of the free parameter $\theta$. As a
consequence, we end up with in total only four cases. These we call in the following Case I to III and Case IV-P1/Case IV-P2.
We first study the mixing patterns analytically for each possible residual symmetry $G_e$ in the charged lepton sector and then
show the results of a $\chi^2$ analysis of these patterns, since they can accommodate the experimental data best. We briefly
comment on a fifth mixing pattern that can fit the data also well apart from the solar mixing angle whose value turns out to be slightly
smaller than the lower $3 \, \sigma$ bound on $\sin^2\theta_{12}$ \cite{nufit}.

\mathversion{bold}
\subsection{$G_e=Z_5$: Case I and Case II}
\mathversion{normal}
\label{sec31}

If we consider as $G_e$ a $Z_5$ symmetry, we find six different categories of tuples $(Q,Z,X)$ with $Q$ being a generator of a $Z_5$ group, taking
into account the mentioned operations in order to relate different tuples $(Q,Z,X)$.
 We can show that for each of these categories we can find a representative tuple
$(Q,Z,X)$ with $Q=T$. Thus, the mixing matrix $U_e$ resulting from the charged lepton sector is the unit matrix, up to permutations of its columns
and unphysical phases. Out of these six representative
tuples four lead to a mixing pattern that we dismiss, because the column that is fixed by group theory is not compatible
with the data at the $3 \, \sigma$ level or better \cite{nufit}. The two remaining representatives,
which we can choose as
\begin{equation}
\label{GeZ5tuple1}
(Q,Z,X) = (T,T^2 S T^3 S T^2, S X_0) \;\;\; \mbox{(Case I)}
\end{equation}
and
\begin{equation}
\label{GeZ5tuple2}
(Q,Z,X) = (T,S T^2 S T, X_0)  \;\;\; \mbox{(Case II)} \; ,
\end{equation}
give rise to a mixing matrix with a column 
whose components have the absolute values
\begin{equation}
\left(
\begin{array}{c}
\sin\varphi \\ \cos\varphi /\sqrt{2}\\ \cos\varphi/\sqrt{2}
\end{array}
\right) \approx \left(
\begin{array}{c}
 0.526\\ 0.602\\ 0.602
\end{array}
\right)\,,
\end{equation}
where we defined for convenience
\begin{equation}
\tan\varphi=1/\phi \; .
\end{equation}
This column can only be identified with the second one of the PMNS mixing matrix, if good agreement with the experimental data should be achieved.
Notice that the ordering of the components in this column as well as its position within the PMNS mixing matrix are not fixed by the approach we are using,
since no constraints on the lepton masses are imposed. 

We first derive the mixing pattern for the tuple in (\ref{GeZ5tuple1}). 
As explained $U_e=1$ and we can take $\Omega_{\footnotesize\mbox{I}}$ to be
\begin{equation}
\label{GeZ5Omega1}
\Omega_{\footnotesize\mbox{I}}= \frac{1}{\sqrt{2}} \, \left(
\begin{array}{ccc}
 \sqrt{2} \, \cos\varphi & -\sqrt{2} \, i \, \sin\varphi & 0\\
 \sin\varphi & i \, \cos\varphi & -1\\
 \sin\varphi &  i \, \cos\varphi & 1
\end{array}
\right) \; ,
\end{equation}
that fulfills (\ref{Omegacon}) for $X$ and $Z$ chosen as in (\ref{GeZ5tuple1}). 
Since $z_1$ and $z_3$ of the (diagonal) combination $\Omega_{\footnotesize\mbox{I}}^\dagger \, Z  \, \Omega_{\footnotesize\mbox{I}}$ are
equal, see (\ref{Omegacon}) for definition, the correct indices $ij$ of the rotation matrix $R_{ij} (\theta)$ in (\ref{formUnu}) are $ij=13$. Thus, the PMNS mixing matrix reads 
\begin{equation}
\label{UPMNS1}
U_{PMNS}= \Omega_{\footnotesize\mbox{I}} \, R_{13} (\theta) \, K_\nu \; .
\end{equation}
We can extract the mixing angles from (\ref{UPMNS1}) in the usual way and find
\begin{eqnarray}\nonumber
&&\sin^2 \theta_{12} = \frac{2}{2 + (3+\sqrt{5}) \, \cos^2 \theta} \;\;\; , \;\;\; \sin^2\theta_{13}= \frac{1}{10} \, \left( 5+\sqrt{5} \right) \, \sin^2 \theta \; ,
\\  \label{anglesCaseI}
&&\sin^2 \theta_{23}= \frac 12 - \frac{\sqrt{2 \, (5+\sqrt{5})} \, \sin 2\theta}{7+\sqrt{5} + (3+\sqrt{5}) \, \cos 2 \theta}  \;\;\; \mbox{(Case I)} \; .
\end{eqnarray}
Furthermore, we can derive the following exact sum rule
among the solar and the reactor mixing angles
\begin{equation}
\label{GeZ5sumrule1}
\sin^2 \theta_{12}= \frac{\sin^2\varphi}{1-\sin^2 \theta_{13}} \approx \frac{0.276}{1-\sin^2 \theta_{13}} \gtrsim 0.276 \; .
\end{equation}
This sum rule can also be directly obtained from $|U_{e2}|^2 = \sin^2 \varphi= 1/(1+\phi^2) \approx 0.276$.
Using for $\sin^2\theta_{13}$ its best fit value $(\sin^2 \theta_{13})^\mathrm{bf}=0.0219$ we find for the solar mixing angle $\sin^2\theta_{12} \approx 0.283$
which is within the $3 \, \sigma$ range, see (\ref{nufitresults}). This value
coincides with the one obtained from a $\chi^2$ analysis, see table \ref{table1}.
The non-trivial lower bound on the solar mixing angle, see (\ref{GeZ5sumrule1}), is also nicely seen in figure \ref{figure1}
(and implicitly also in the $\sin^2\theta_{13}$-$\sin^2 \theta_{12}$ plane in figure \ref{figure2}).
For the atmospheric mixing angle a simple approximate relation to the reactor mixing angle is given by
\begin{equation}
\!\!\!\!\sin^2\theta_{23} \approx \frac 12 \, \left( 1\pm (1-\sqrt{5}) \, \sin\theta_{13} \right) 
\approx 0.5 \mp 0.618 \, \sin\theta_{13}
\approx  \left\{ \begin{array}{c} 0.409 \;\; \mbox{for} \;\; \theta < \pi/2 \\
0.591 \;\; \mbox{for} \;\; \theta > \pi/2  \end{array} \right. \, ,
\end{equation}
if we use again the best fit value $(\sin^2 \theta_{13})^\mathrm{bf}=0.0219$. 
The first subleading term arises at $\sin^3\theta_{13}$ and can thus be safely neglected in this estimate. 
We can clearly see having a look at the different symbols in figures \ref{figure1} and \ref{figure2} that values of $\theta \leq \pi/2$ 
lead to $\sin^2\theta_{23} \leq 1/2$, while larger values, $\pi/2 \leq \theta \leq \pi$, entail $\sin^2\theta_{23} \geq 1/2$. This is also confirmed 
by the two different `best fitting' values  $\theta_{\footnotesize\mbox{bf}} \approx 0.174$ and $\theta_{\footnotesize\mbox{bf}} \approx 2.967$ 
 that are obtained for NO and IO, respectively, see table \ref{table1}.

Note that the formulae for the mixing angles $\theta_{12}$ and $\theta_{13}$ remain invariant, if we replace $\theta$ by $\pi-\theta$, 
while the relative sign in the expression for $\sin^2 \theta_{23}$ changes. The same effect can be achieved, 
if we consider the PMNS mixing matrix in (\ref{UPMNS1}) with second and third rows exchanged.

As one can check, all CP invariants $J_{CP}$, $I_1$ and $I_2$ vanish exactly and thus an accidental CP symmetry must be present.
Indeed, there is one, namely
\begin{equation}
\label{Y1}
Y= \left( \begin{array}{ccc}
 1 & 0 & 0\\
 0 & 1 & 0\\
 0 & 0 & 1
\end{array}
\right) \,,
\end{equation}
that fulfills the conditions in (\ref{CPacc1})-(\ref{CPacc2}) for the tuple $(Q,Z,X)$ shown in (\ref{GeZ5tuple1}).
The vanishing of $\sin\delta$ can also be confirmed by verifying that the condition presented in \cite{twoCPs}
is fulfilled.

In a similar manner we can study the lepton mixing that can be obtained for the second tuple $(Q,Z,X)$, the one in (\ref{GeZ5tuple2}).
The form of the matrix $\Omega_{\footnotesize\mbox{II}}$ that fulfills (\ref{Omegacon}) for $Z=S T^2 S T$ and $X=X_0$ can be chosen as
\begin{equation}
\label{GeZ5Omega2}
\Omega_{\footnotesize\mbox{II}}= \frac{1}{\sqrt{2}} \, \left(
\begin{array}{ccc}
 -\sqrt{2} \, \cos\varphi & -\sqrt{2} \, \sin\varphi & 0\\
 -\, e^{-3 \, i \, \Phi}\, \sin\varphi & e^{-3\, i\, \Phi} \, \cos\varphi & -e^{-7 \, i \, \Phi/4}\\
 - e^{-2 \, i \, \Phi} \, \sin\varphi &  e^{-2 \, i \, \Phi} \, \cos\varphi & e^{- 3 \, i \, \Phi/4}
\end{array}
\right) \; .
\end{equation}
Like in the preceding case, also here the necessary rotation $R_{ij} (\theta)$ is in the (13)-plane. Thus, taking into account that $U_e$ is trivial,
the PMNS mixing matrix is of the form (up to permutations of rows and columns and unphysical phases)
\begin{equation}
\label{UPMNS2}
U_{PMNS}= \Omega_{\footnotesize\mbox{II}} \, R_{13} (\theta) \, K_\nu \; .
\end{equation}
The predictions for the solar and the reactor mixing angles are the same as for the first tuple, namely
\begin{equation}
\sin^2 \theta_{12} = \frac{2}{2 + (3+\sqrt{5}) \, \cos^2 \theta} \;\;\; , \;\;\; \sin^2\theta_{13}= \frac{1}{10} \, \left( 5+\sqrt{5} \right) \, \sin^2 \theta  \;\;\; \mbox{(Case II)} \; .
\end{equation}
Thus, also the sum rule and the estimate given in and below (\ref{GeZ5sumrule1}) hold. In figure \ref{figure1} the non-trivial
lower bound on $\sin^2 \theta_{12}$ is visible also for Case II.
The results for the atmospheric mixing angle and the Dirac phase instead are different, since both of them are predicted to be maximal
\begin{equation}
\sin^2 \theta_{23} = \frac 12 \;\;\; \mbox{and} \;\;\; |\sin\delta|=1  \;\;\; \mbox{(Case II)} \; .
\end{equation}
The sign of $\sin\delta$ depends on whether $\theta \lessgtr \pi/2$.
The corresponding Jarlskog invariant $J_{CP}$ reads\footnote{For $\sin 2\theta=0$ the Jarlskog invariant vanishes. If this happens, one of the mixing 
angles becomes either $0$ or $\pi/2$ and thus the Dirac phase $\delta$
becomes unphysical. Clearly, these values of $\theta$ are highly disfavored by experimental data.}
\begin{equation}
\label{JCP2}
J_{CP}=-\frac{1}{20 \, \sqrt{2}} \, \sqrt{5+\sqrt{5}} \, \sin 2 \theta \; . 
\end{equation}
Like in the first case also in this case the Majorana phases are trivial. We can check that the conditions found in \cite{twoCPs} 
for $\cos\delta=0$ and $\theta_{23}=\pi/4$ are fulfilled in the case at hand. Furthermore, we can verify with the help of the formulae given in \cite{twoCPs}
that both Majorana phases must be trivial.

We notice that the replacement of the parameter $\theta$ by $\pi-\theta$ does not change the form of the mixing angles, while the sign of $J_{CP}$ and, consequently, of $\sin\delta$
changes. The very same result is achieved, if we exchange the second and third rows of the mixing matrix in (\ref{UPMNS2}). Consequently, we expect to find
two best fitting values of the parameter $\theta$. This is confirmed by the results of the $\chi^2$ analysis in table \ref{table1}. 
As expected, the sum of these two best fitting values equals $\pi$.

One of the categories initially dismissed by the criterion that the absolute values of the group theoretically fixed column of the PMNS mixing matrix
should agree at the $3 \, \sigma$ level or better with the values given in \cite{nufit} might still be interesting in a concrete model
in which (small) corrections can lead to the agreement with experimental data. A representative of this case is the tuple $(Q,Z,X)=(T,S,X_0)$. 
The absolute values of the elements of the fixed column are $\left( \cos\varphi, \sin\varphi/\sqrt{2}, \sin\varphi/\sqrt{2}\right)^T \approx 
\left( 0.851, 0.372, 0.372 \right)^T$. Thus, this column can be identified with the first one of the PMNS mixing matrix. This pattern fails to 
describe the data well without corrections mainly because of the tight relation between the solar and the reactor mixing angle that can be derived.
We find $\sin^2\theta_{12}=1-\frac{5+\sqrt{5}}{10 \, (1-\sin^2\theta_{13})} \lesssim 0.276$ that leads for $(\sin^2 \theta_{13})^{\mathrm{bf}}=0.0219$ to a too small solar mixing angle 
$\sin^2 \theta_{12} \approx 0.260$. At the same time, the atmospheric mixing angle is maximal. The Dirac phase is also maximal, whereas both Majorana phases are trivial. 
So, this case shows strong similarities to Case II with the representative tuple shown in (\ref{GeZ5tuple2}).

\mathversion{bold}
\subsection{$G_e=Z_3$: Case III}
\mathversion{normal}
\label{sec32}

If we do the same analysis for the case $G_e=Z_3$, we find eight categories of tuples $(Q,Z,X)$.
We can always fix $Q= T^2 S T^2$. In this case $U_e$ is not trivial anymore and is of the form
\begin{equation}
U_e= \left( \begin{array}{ccc}
  -\sqrt{\frac{7+3 \sqrt{5}}{3 (5+\sqrt{5})}} & -\sqrt{\frac{(5-\sqrt{5})}{15}} & \frac{2}{\sqrt{3 (5+\sqrt{5})}}\\
  \frac{2}{\sqrt{3 (5 +\sqrt{5})}} & -\frac 12 - \frac{1}{30} \, \sqrt{75+30 \sqrt{5}}& -\frac 12 + \frac{1}{30} \, \sqrt{75+30 \sqrt{5}}\\
  \frac{2}{\sqrt{3 (5 +\sqrt{5})}} & \frac 12 - \frac{1}{30} \, \sqrt{75+30 \sqrt{5}}& \frac 12 + \frac{1}{30} \, \sqrt{75+30 \sqrt{5}}
\end{array} \right) \; .
\end{equation}
Using the representatives of the eight different categories we see that indeed only one of these can lead to a mixing that is compatible 
with experimental data. The column that is fixed by group theory is in this case TM \cite{TM}
\begin{equation}
\frac{1}{\sqrt{3}} \left(
\begin{array}{c}
 1\\ 1\\ 1
\end{array}
\right) \approx \left(
\begin{array}{c}
 0.577\\ 0.577\\ 0.577
\end{array}
\right)
\end{equation}
and has therefore to be identified with the second column of the PMNS mixing matrix.
Immediately, we know that the solar mixing angle has a lower bound, $\sin^2 \theta_{12} \gtrsim 1/3$, see the sum rule in (\ref{GeZ3sumrule}),
the result of the $\chi^2$ analysis in table \ref{table1} and the lower bound in figure \ref{figure1}.
We use as representative the tuple
\begin{equation}
\label{GeZ3tuple}
(Q,Z,X)= (T^2 S T^2, S T^2 S T^3 S, X_0)  \;\;\; \mbox{(Case III)} \; .
\end{equation}
A possible admitted form of the matrix $\Omega$ is
\begin{equation}
\Omega_{\footnotesize\mbox{III}}=\frac{1}{\sqrt{2}} \, \left( \begin{array}{ccc}
\sqrt{2} \, \cos \varphi& 0& -\sqrt{2} \, \sin\varphi\\
e^{i \, \Phi} \, \sin\varphi & e^{9 \, i \, \Phi/4} & e^{i \, \Phi} \, \cos\varphi \\
e^{-i \, \Phi} \, \sin\varphi & e^{11 \, i \, \Phi/4} & e^{-i \, \Phi} \, \cos\varphi 
\end{array}
\right) \; .
\end{equation}
As can be seen, the form of the matrix $\Omega$ is quite similar to the ones used in the other cases, see  (\ref{GeZ5Omega1}) and (\ref{GeZ5Omega2}).
The matrix $U_\nu$ is composed as follows
\begin{equation}
\label{Unu3}
U_\nu= \Omega_{\footnotesize\mbox{III}} \, R_{13} (\theta) \, K_\nu \; ,
\end{equation}
since, as in the cases above, $z_1$ and $z_3$ of the matrix combination $\Omega^\dagger_{\footnotesize\mbox{III}} Z \Omega_{\footnotesize\mbox{III}}$ are equal. The PMNS mixing matrix
is then given by
\begin{equation}
\label{UPMNS3}
U_{PMNS}= U_e^\dagger \, \Omega_{\footnotesize\mbox{III}} \, R_{13} (\theta) \, K_\nu \; .
\end{equation}
We can extract the following results for the solar and the reactor mixing angles
\begin{equation}
\sin^2 \theta_{12} = \frac{1}{2+\sin2\theta} \;\;\; \mbox{and} \;\;\; \sin^2 \theta_{13}= \frac 13 \, \left( 1-\sin2\theta \right)  \;\;\; \mbox{(Case III)}
\end{equation}
that fulfill the exact -- and well-known -- sum rule \cite{TM}
\begin{equation}
\label{GeZ3sumrule}
\sin^2 \theta_{12} = \frac{1}{3 \, (1-\sin^2 \theta_{13})} \gtrsim \frac 13 \; .
\end{equation}
 If we use $(\sin^2 \theta_{13})^{\mathrm{bf}}=0.0219$ as best fit value for $\sin^2 \theta_{13}$, we arrive at $\sin^2 \theta_{12} \approx 0.341$ which is the value that is also
obtained in our $\chi^2$ analysis, see table \ref{table1}.
The atmospheric mixing angle as well as the Dirac phase are, as in Case II, maximal
\begin{equation}
\sin^2 \theta_{23} = \frac 12 \;\;\; \mbox{and} \;\;\; |\sin\delta|=1  \;\;\; \mbox{(Case III)} \; .
\end{equation}
Again, the actual sign of $\sin\delta$ depends on the choice of $\theta$.
The form of the Jarlskog invariant is
\begin{equation}
\label{JCP3}
J_{CP}= \frac{\cos 2 \theta}{6 \, \sqrt{3}} \; .
\end{equation}
If $\cos 2\theta$ vanishes, $J_{CP}$ equals $0$. Since at the same time one mixing angle becomes $0$ or $\pi/2$, the Dirac phase $\delta$ turns out
to be unphysical for $\cos2\theta=0$.
As happened for Case I and II, also here the mixing pattern that is well compatible with experimental data gives rise to trivial Majorana phases.

We note that the formulae of the mixing angles remain invariant, if we replace $\theta$ with $\pi/2-\theta$. Thus, we expect
(at least) two best fit solutions. This expectation is confirmed by our $\chi^2$ analysis, see table \ref{table1}.
Replacing $\theta$ with $\pi/2-\theta$ leads, at the same time, to an additional sign for $J_{CP}$ and thus $\sin\delta$.
Similarly, the exchange of the second and third rows of the PMNS mixing matrix in (\ref{UPMNS3}) does not alter the results for the mixing angles, but changes the sign of the Jarlskog invariant and thus of $\sin\delta$.

\mathversion{bold}
\subsection{$G_e=Z_2 \times Z_2$: Case IV-P1 and Case IV-P2}
\mathversion{normal}
\label{sec33}

For the remaining possibility $G_e=Z_2 \times Z_2$ we find that all admitted combinations 
of $Q_1$, $Q_2$ with $Z$ and $X$ describing the residual symmetry $G_\nu=Z_2 \times CP$ can be classified in four 
different categories of tuples $(\left\{Q_1, Q_2 \right\},Z,X)$. Thus, it is sufficient to calculate the mixing pattern for one representative of each category.
Note that we can always choose a representative for which $G_e=K_1$, see (\ref{Kgens}), i.e. $Q_1$ and $Q_2$
can be chosen as $Q_1=S$ and $Q_2=T^2 S T^3 S T^2$. So, the form of the matrix $U_e$ is
\begin{equation}
\label{GeZ2Z2Ue}
U_e=\frac{1}{\sqrt{2}} \, \left( \begin{array}{ccc}
 \sqrt{2} \, \cos\varphi & 0 & -\sqrt{2} \, \sin\varphi\\
 \sin\varphi & -1 & \cos\varphi\\
 \sin\varphi & 1 & \cos\varphi
\end{array}
\right) \; .
\end{equation}
It turns out that only one category of tuples  is capable of accommodating the experimental values of the mixing angles well for a particular choice of the parameter
$\theta$, while the other three ones fail to do so. In particular, two out of these three lead to patterns with only one non-vanishing mixing angle,
since the generators $Q_i$ are diagonalized by the same matrix as the $Z_2$ generator $Z$. A representative of the category that
allows for good agreement with the data is
\begin{equation}
\label{GeZ2Z2tuple}
(\left\{ Q_1, Q_2 \right\}, Z, X)=\left( \left\{ S , T^2 S T^3 S T^2 \right\} , S T^2 S T , X_0 \right)  \;\;\; \mbox{(Case IV)} \; . 
\end{equation}
We can check that the column that does not depend on the free parameter $\theta$ has components with absolute values of the form
\begin{equation}
\label{GeZ2Z2vector}
\frac 12 \, \left(
\begin{array}{c}
 \phi \\ 1 \\ 1/\phi
\end{array}
\right) \approx \left(
\begin{array}{c}
 0.809 \\ 0.5 \\ 0.309 
\end{array}
\right) \; .
\end{equation}
Thus, this column can be identified with the first one of the PMNS mixing matrix. We call this situation Case IV-P1.
We note that we can exchange the second and third
components of the vector in (\ref{GeZ2Z2vector}), i.e. we can exchange the second and third rows of the resulting PMNS mixing matrix, and also obtain good agreement with the 
experimental data. This situation is denoted by Case IV-P2 in the following. The crucial change occurs in the predicted 
value of the atmospheric mixing angle, see (\ref{GeZ2Z2sinth23}) and (\ref{GeZ2Z2sinth23_2}). 

A unitary matrix $\Omega$ that fulfills the conditions in (\ref{Omegacon}) for $Z=S T^2 S T$ and $X=X_0$ is 
\begin{equation}
\!\!\!\!\!\!\!\!\Omega_{\footnotesize\mbox{IV}}= \frac{1}{\sqrt{2}} \, \left( \begin{array}{ccc}
\sqrt{2}\, \sin\varphi & -\sqrt{2} \, \cos^2\varphi & \sqrt{2} \, \cos\varphi \, \sin\varphi\\
 e^{- i \, \Phi/2} \, \cos\varphi & e^{-i \, \Phi/2} \, \left( e^{15 \, i \, \Phi/4} +\cos\varphi \right) \, \sin\varphi & e^{-i \, \Phi/2} \, \left( e^{15 \, i \, \Phi/4} \, \cos\varphi - \sin^2 \varphi \right) \\
-e^{- 2 \, i \, \Phi} \, \cos\varphi & e^{-2 \, i \, \Phi} \, \left( e^{15 \, i \, \Phi/4} -\cos\varphi \right) \, \sin\varphi & e^{-2 \, i \, \Phi} \, \left( e^{15 \, i \, \Phi/4} \, \cos\varphi + \sin^2 \varphi \right)
\end{array}
\right) \; .
\end{equation}
Since the diagonal matrix $\Omega^\dagger_{\footnotesize\mbox{IV}} \, Z \, \Omega_{\footnotesize\mbox{IV}}$ 
reveals equal $z_2$ and $z_3$ in the convention of (\ref{Omegacon}), the form of the
 neutrino mixing matrix is given by, up to permutations of columns and unphysical phases,
\begin{equation}
U_\nu= \Omega_{\footnotesize\mbox{IV}} \, R_{23} (\theta) \, K_\nu \, .
\end{equation}
Taking into account the contribution to leptonic mixing coming from the charged lepton sector that is encoded in the matrix $U_e$ in (\ref{GeZ2Z2Ue}),
the PMNS mixing matrix is of the form, up to permutations of rows and columns and unphysical phases,
\begin{equation}
U_{PMNS}= U_e^\dagger \, \Omega_{\footnotesize\mbox{IV}} \, R_{23} (\theta) \, K_\nu \, .
\end{equation}
We find for the mixing angles the following expressions
\begin{eqnarray}\label{GeZ2Z2sinth1213}
&&\sin^2 \theta_{12}= \frac{(5-2 \, \sqrt{5}) \, \cos^2 \theta}{1+ (5-2 \, \sqrt{5}) \, \cos^2 \theta} \;\;\; , \;\;\;
\sin^2 \theta_{13}= \frac 18 \, \left( 5-\sqrt{5} \right) \, \sin^2 \theta  \; \mbox{(Case IV)} 
\\ \label{GeZ2Z2sinth23}
&&\sin^2 \theta_{23}= \frac{3 \, (5-\sqrt{5}) + (9-5 \, \sqrt{5}) \, \cos 2 \theta + 8 \, \sin 2 \theta}{25-3 \, \sqrt{5} + 5 \, (3-\sqrt{5}) \, \cos 2 \theta}  \;\;\; \mbox{(Case IV-P1)} \; .
\end{eqnarray}
For this case, as mentioned above, the absolute values of the elements of the first column of the PMNS mixing matrix are ordered in the same way as in (\ref{GeZ2Z2vector}).
From (\ref{GeZ2Z2sinth1213}) we can derive an exact sum rule relating the solar mixing angle to the reactor one
\begin{equation}
\label{GeZ2Z2sumrule1}
\sin^2 \theta_{12} = 1 - \frac{3+\sqrt{5}}{8 \, (1-\sin^2 \theta_{13})} \approx 0.331 \; ,
\end{equation}
if we insert the experimental best fit value $(\sin^2 \theta_{13})^{\mathrm{bf}}=0.0219$. 
We note that in this case a non-trivial upper bound on $\sin^2 \theta_{12}$ exists, namely $\sin^2 \theta_{12}\lesssim 1-\phi^2/4 \approx 0.345$.
This can also be directly derived from the constraint that $|U_{e1}|^2=\phi^2/4= (3+\sqrt{5})/8 \approx 0.655$.
This bound is marked with a (dashed red) vertical line in figure \ref{figure1}.
 Furthermore we can obtain an approximate
relation of the atmospheric mixing angle and $\sin\theta_{13}$
\begin{equation}
\sin^2 \theta_{23} \approx \frac{1}{10} \, \left( 5-\sqrt{5} \right) + \frac 25 \, \sqrt{5+2 \, \sqrt{5}} \, \sin \theta_{13} \approx 0.276 + 1.23 \, \sin \theta_{13} \; ,
\end{equation}
using that $\theta$ lies in the interval $[0, \pi/2]$ (and thus $\cos\theta=\sqrt{1-\sin^2\theta}$).\footnote{\label{f1} For $\pi/2 \leq \theta \leq \pi$ we can derive a similar relation that shows, however, that the measured values of the atmospheric and the reactor mixing angle cannot be 
accommodated well at the same time. }
Subleading corrections are of order $\sin^2 \theta_{13}$ at maximum. For $(\sin^2 \theta_{13})^{\mathrm{bf}}=0.0219$,
we find
\begin{equation}
\sin^2 \theta_{23} \approx 0.459 \; .
\end{equation}
This estimate is consistent with the result of our $\chi^2$ analysis, see table \ref{table1}.

If we permute the second and the third rows, the form of the reactor as well as of the solar mixing angle is the same, while the atmospheric
one turns out to be $1-\sin^2 \theta_{23}$, i.e. here the atmospheric mixing angle reads
\begin{equation}
\label{GeZ2Z2sinth23_2}
\sin^2 \theta_{23}=1-  \frac{3 \, (5-\sqrt{5}) + (9-5 \, \sqrt{5}) \, \cos 2 \theta + 8 \, \sin 2 \theta}{25-3 \, \sqrt{5} + 5 \, (3-\sqrt{5}) \, \cos 2 \theta} \;\;\; \mbox{(Case IV-P2)} \; .
\end{equation}
So, in this case the approximate sum rule relating $\theta_{23}$ and $\theta_{13}$ is given by 
\begin{equation}
\sin^2 \theta_{23} \approx \frac{1}{10} \, \left( 5+\sqrt{5} \right) - \frac 25 \, \sqrt{5+2 \, \sqrt{5}} \, \sin \theta_{13} \approx 0.724 - 1.23 \, \sin \theta_{13} \; .
\end{equation}
Again, we assume $0 \leq \theta \leq \pi/2$ (and for $\theta$ lying in the interval $[\pi/2, \pi]$ see footnote \ref{f1}). For $(\sin^2 \theta_{13})^{\mathrm{bf}}=0.0219$ we get
\begin{equation}
\sin^2 \theta_{23} \approx 0.541 
\end{equation}
which is consistent with the value obtained from the $\chi^2$ fit, see table \ref{table1}.

All CP phases are trivial. This points towards an accidental CP symmetry in the theory. This is clear, since the CP transformation $X=X_0$
is not only present in the neutrino sector, but also -- as one can check explicitly -- in the charged lepton one. 

Note that there is no evident symmetry as regards the parameter $\theta$ in the formulae of the mixing angles: while $\theta_{12}$ and $\theta_{13}$
remain invariant, if we replace the parameter $\theta$ with $\pi-\theta$, this is not the case for the atmospheric mixing angle and thus we expect
in general only one value of $\theta$ for which the mixing angles can be accommodated best. This is confirmed in our numerical analysis, see table \ref{table1}.

\subsection{Numerical discussion}
\label{sec34}

In the following we present our results of a $\chi^2$ analysis for the different cases, Case I through IV-P2. The 
$\chi^2$ function is defined in the usual way
\begin{eqnarray}
\label{chisqtot}
&&\chi^2 = \chi^2_{12} + \chi^2_{13} + \chi^2_{23}
\\ \label{chisqsingle}
\mbox{with}\;\;\;&&\chi^2_{ij} = \left(\frac{\sin^2 \theta_{ij} - (\sin^2 \theta_{ij})^{\mathrm{bf}}}{\sigma_{ij}} \right)^2 \;\;\; \mbox{for} \;\;\; ij=12, 13, 23 \; .
\end{eqnarray}
$\sin^2 \theta_{ij}$ are the mixing angles derived in the different cases, e.g. see (\ref{anglesCaseI}) for Case I, that depend on the continuous parameter $\theta$, ranging from $0$ to $\pi$,
$\left(\sin^2 \theta_{ij}\right)^{\mathrm{bf}}$ are the best fit values and 
$\sigma_{ij}$ the $1 \, \sigma$ errors reported in (\ref{nufitresults}). Note that these errors also depend on whether $\sin^2 \theta_{ij}$ is larger or smaller than the best fit
 value. Since the global fit results for the mixing angles (slightly) differ for the case of NO or IO, we consider these two separately
 and calculate for all patterns the $\chi^2$ function $\chi^2_{\mathrm{NO}}$ under the assumption of NO being realized in nature and 
 $\chi^2_{\mathrm{IO}}$ for IO. In particular, in doing so we do not take into account the fact that NO is slightly disfavored by $\Delta \chi^2=0.97$ compared to IO \cite{nufit}.
 A mixing pattern is considered to agree reasonably well with the experimental
 data, if $\chi^2 \lesssim 27$ and all mixing angles $\sin^2 \theta_{ij}$ are within the $3\, \sigma$ intervals 
 given in (\ref{nufitresults}).\footnote{
The upper limit on $\chi^2$, $\chi^2\lesssim 27$, is chosen, since it results from summing three
$3 \, \sigma$ gaussian errors for one degree of freedom.
}
The $\chi^2$ functions $\chi^2_{\mathrm{NO}}$ and $\chi^2_{\mathrm{IO}}$ are minimized at the best fitting point(s) $\theta=\theta_{\mathrm{bf}}$ and we only report the global minimum/a in table \ref{table1} for each case.\footnote{
We have tested the validity of our $\chi^2$ analysis by constructing a likelihood function to fit the various cases that uses the one-dimensional $\chi^2$ 
projections provided in \cite{nufit}. These results are consistent with those in table \ref{table1}, up to the fact that the roles of the local and the global 
minimum in Case I become exchanged for NO. However, the difference between these two minima turns out to be statistically insignificant.
} Since the indication of a preferred value of the Dirac phase $\delta$ coming 
 from global fit analyses is rather weak \cite{nufit}, i.e. below the $3\, \sigma$ significance, we do not include any information on $\delta$ in the $\chi^2$ function in (\ref{chisqtot}).

\begin{table}[t!]
\hspace{-0.6in}
\begin{tabular}{c}
$
\begin{array}{|l|c|c|c|c|c|c|c|c|c|c|}
\hline
 &  \multicolumn{2}{c|}{\text{Case I}} &   \multicolumn{2}{c|}{\text{Case II}} &   \multicolumn{2}{c|}{\text{Case III}} &   \multicolumn{2}{c|}{\text{Case IV-P1}} &  \multicolumn{2}{c|}{\text{Case IV-P2}} \\
\cline{2-11}\rule[0.05in]{0cm}{0cm}
\text{\footnotesize $(Q_i,Z,X)$} &  \multicolumn{2}{c|}{\text{\footnotesize $\left(T, T^2 S T^3 S T^2, S X_0 \right)$}} &   \multicolumn{2}{c|}{\text{\footnotesize$\left( T, S T^2 S T, X_0\right)$}} &   \multicolumn{2}{c|}{\text{\footnotesize$\left( T^2 S T^2, S T^2 S T^3 S, X_0\right)$}} &  
 \multicolumn{4}{c|}{\text{\footnotesize$\left( \left\{ S , T^2 S T^3 S T^2 \right\} , S T^2 S T , X_0 \right)$}} \\[0.02in]
 \cline{2-11}\normalsize 
 &\rule[0.15in]{0cm}{0cm} \text{NO}&\text{IO} & \text{NO}&\text{IO} & \text{NO}&\text{IO} & \text{NO}&\text{IO} & \text{NO}&\text{IO}\\  
 \hline
\rule[0.15in]{0cm}{0cm}\chi^2_{\footnotesize\mbox{min}}  & 5.64 & 3.46 & 4.04 & 7.74 & 8.84 & 12.56 & 4.48 & 11.80 & 6.19 & 6.43 \\
\theta_{\footnotesize\mbox{bf}} &0.174 & 2.967 &  \multicolumn{2}{c|}{\left\{ \begin{array}{c} 0.175 \\ 2.967
\end{array}\right.} & \left\{ \begin{array}{c} 0.604 \\ 0.967\end{array}\right. & \left\{ \begin{array}{c} 0.603 \\ 0.967
\end{array}\right.  & 0.254 & 0.258 & 0.255 & 0.254 \\[0.1in]
 \hline
\rule[0.15in]{0cm}{0cm}\sin^2\theta_{12} & 0.283 & 0.283  & 0.283 & 0.283 & 0.341 & 0.341 & 0.331 & 0.330 & 0.331 & 0.331 \\
\sin^2\theta_{13}  & 0.0217 & 0.0219 & 0.0218 & 0.0220 & 0.0217 & 0.0218 & 0.0219 & 0.0225 & 0.0220 & 0.0218 \\
\sin^2\theta_{23}  & 0.408 & 0.592 & \multicolumn{2}{c|}{0.5}  & \multicolumn{2}{c|}{0.5}  & 0.475 & 0.478 & 0.524 & 0.525\\
\hline
J_{CP} & 0 & 0 & \mp 0.0325 & \mp 0.0326 & \pm 0.0342 & \pm 0.0342 & 0 & 0 & 0 & 0\\
\sin\delta & 0 & 0 &  \multicolumn{2}{c|}{\mp 1}   &  \multicolumn{2}{c|}{\pm 1}  & 0 & 0 & 0 & 0\\
 \hline
\end{array}
$
\end{tabular}
\caption{{\small {\it Results of $\chi^2$ analysis for Case I through Case IV-P2, displayed separately for the assumption of NO and IO. We remind the reader of the 
representative tuples $(Q_i,Z,X)$ that we have chosen for the different cases.
$\chi^2_{\footnotesize\mbox{min}}$ is the smallest value of $\chi^2$ that can be obtained for a particular mixing pattern at the
best fitting value(s) $\theta_{\footnotesize\mbox{bf}}$. We mention for each pattern and neutrino mass ordering only the lowest 
value of $\chi^2_{\footnotesize\mbox{min}}$ that can be achieved. All values of $\sin^2\theta_{ij}$ are obtained at the given $\theta_{\footnotesize\mbox{bf}}$.
For Case II and Case III that predict maximal atmospheric mixing and a maximal Dirac
phase $\delta$, the sign of the Jarlskog invariant $J_{CP}$ (and of $\sin\delta$) depends on the chosen value of $\theta_{\footnotesize\mbox{bf}}$
and upper (lower) signs correspond to the smaller (larger) value of $\theta_{\footnotesize\mbox{bf}}$.
Majorana phases are trivial in all cases, $\sin\alpha=0$ and $\sin\beta=0$.}}}
\label{table1}
\end{table}
%
%

\begin{figure}[t!]
\begin{center}
\begin{tabular}{cc}
\includegraphics[width=0.5\textwidth]{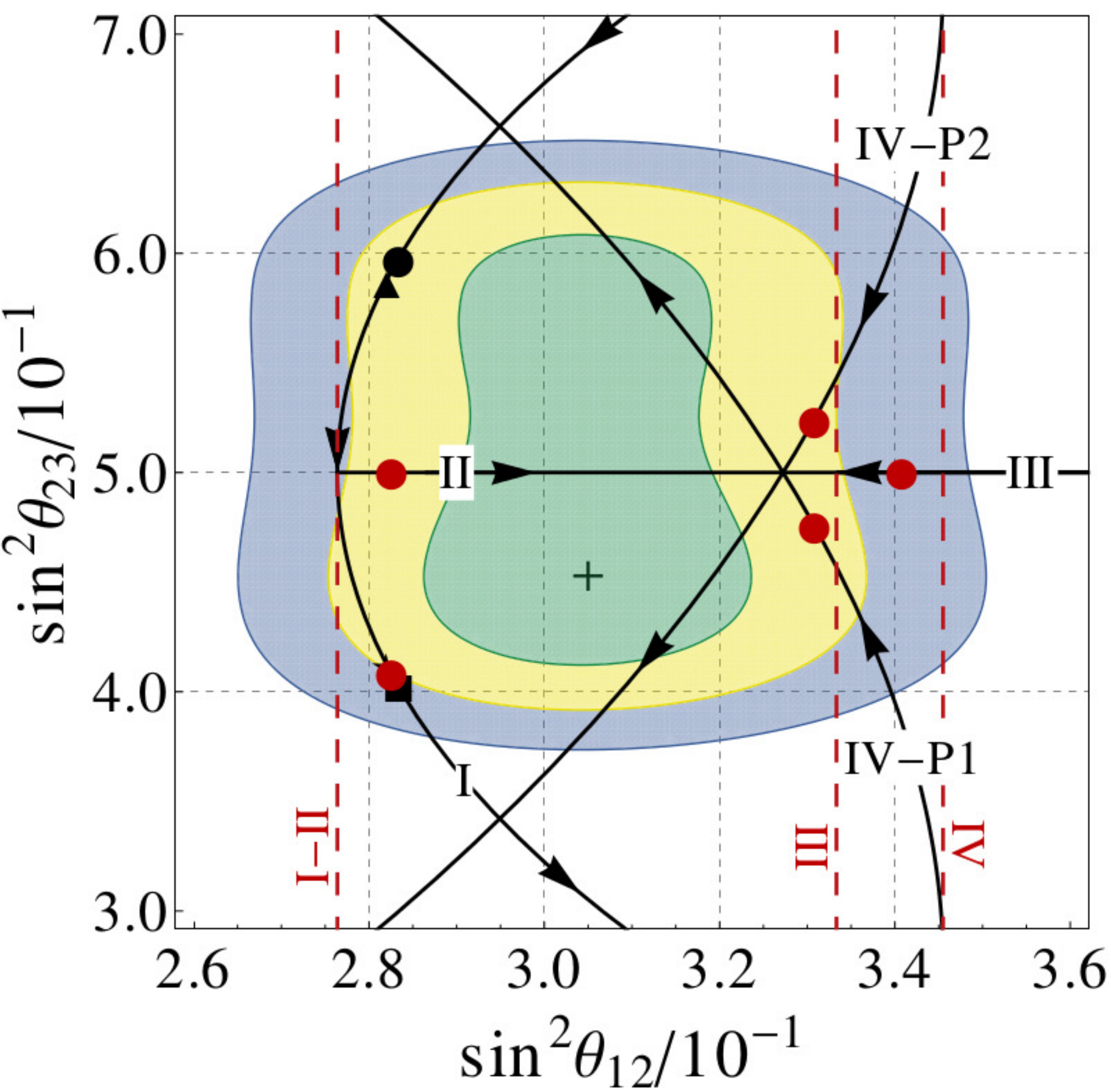} &
\includegraphics[width=0.5\textwidth]{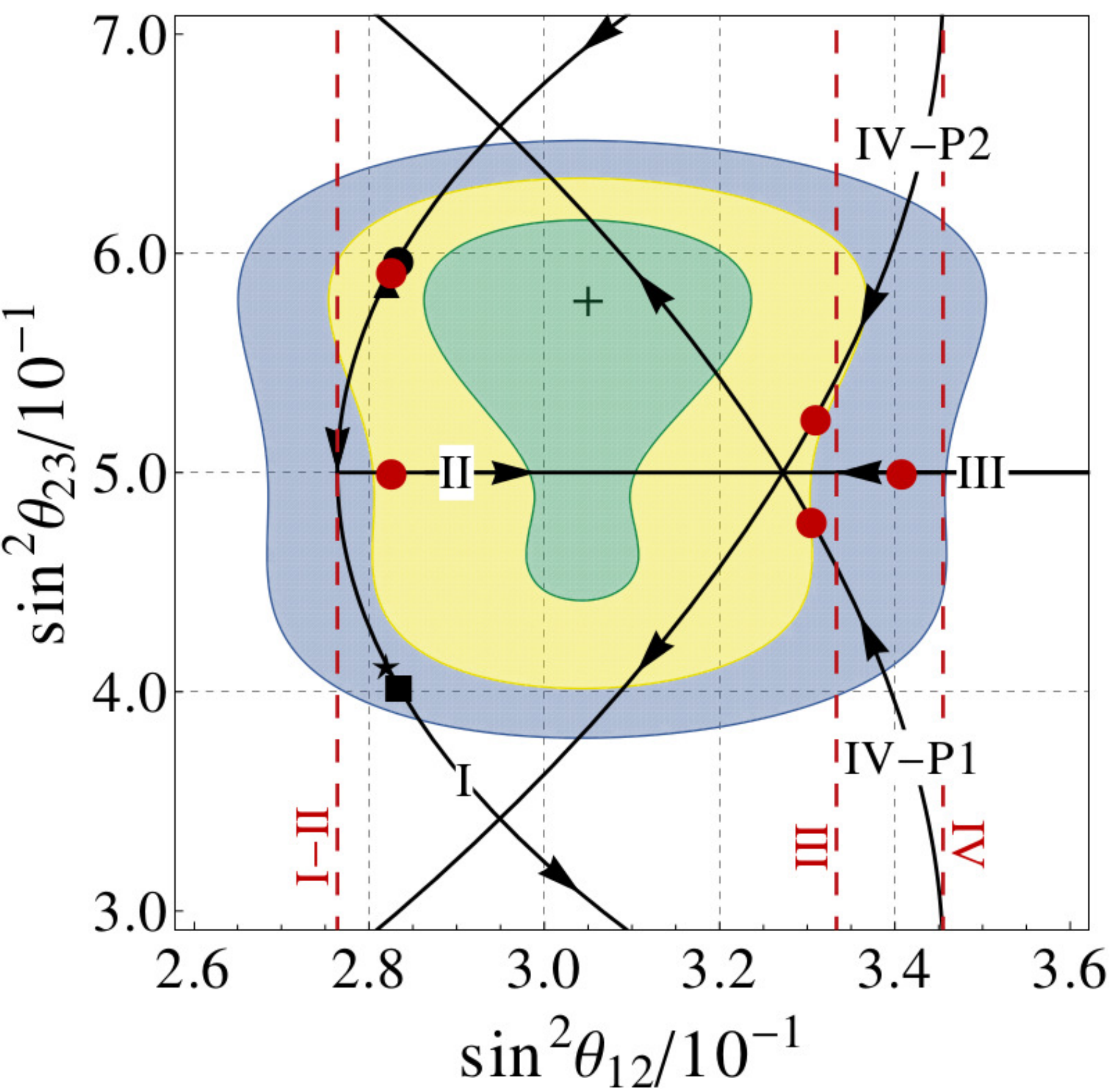}
\end{tabular}
\caption{\small {\it 
Results for the mixing angles $\sin^{2}\theta_{23}$ and $\sin^2\theta_{12}$ for Case I to III and Case IV-P1 and IV-P2. 
The plot on the left side assumes NO, while the one on the right IO. The curves of the
various cases are superimposed to the experimentally preferred 
$1 \, \sigma$ (green), $2 \, \sigma$ (yellow) and $3 \, \sigma$ (blue) areas
at two degrees of freedom adapted from \cite{nufit}. In drawing these areas we have subtracted $\Delta \chi^2 =0.97$ in the case of NO  so that the minimal value of $\Delta \chi^2$
is zero for both mass orderings. The experimental best fit value is indicated with a cross. The best fitting value of $\theta$ is indicated on 
each curve with a red dot.
For clarity we mark on the curve belonging to Case I  some particular values of the parameter $\theta$ with different symbols in black:
star for $\theta=\pi/19$, square for $\theta=\pi/17$, dot for $\theta=16\, \pi/17$ and triangle for $\theta=18\, \pi/19$.
In addition, an arrow indicates on each curve the direction
of increasing $\theta$ in the interval $0 \leq \theta \leq \theta_{\footnotesize\mbox{bf} (,1)}$ (the arrow always belongs to the closest label I, II, etc.). 
Note that the curves belonging to Case II and Case III partly overlap. Dashed red vertical lines
indicate the non-trivial lower (upper) bound on the solar mixing angle in Case I, II and III (Case IV-P1 and IV-P2).}}
\label{figure1}
\end{center}
\end{figure}

Our findings for the different cases are summarized in table \ref{table1}. As one can see, these results agree well with our analytical estimates and observations made in subsections
\ref{sec31}-\ref{sec33}. In particular, the sum of the two best fitting values of $\theta$ for NO and IO in Case I, $\theta_{\footnotesize\mbox{bf, NO}}$
and $\theta_{\footnotesize\mbox{bf, IO}}$, approximately equals $\pi$, since the formulae for the mixing angles
$\theta_{12}$ and $\theta_{13}$ are invariant under the transformation $\theta \; \rightarrow \; \pi-\theta$, while $\sin^2\theta_{23}$ turns into $\cos^2\theta_{23}$. 
Similarly, the sum of the two best fitting points $\theta_{\footnotesize\mbox{bf}, 1}$ and $\theta_{\footnotesize\mbox{bf}, 2}$ ((almost) the same for NO and IO)
equals $\pi$ in Case II. Also related to the symmetry properties of the formulae for the solar and the reactor mixing angles is the observation in Case III that 
 the two best fitting points (for NO and IO), $\theta_{\footnotesize\mbox{bf}, 1}$ and $\theta_{\footnotesize\mbox{bf}, 2}$,
sum up to $\pi/2$. Case IV does not reveal such a symmetry in the parameter $\theta$ and thus we discuss in this case the results corresponding to two 
different permutations, Case IV-P1 and Case IV-P2, that are related by the exchange of the second and third rows of the PMNS mixing matrix. 
This allows us to accommodate $\sin^2\theta_{23}< 1/2$ as well as $\sin^2\theta_{23}>1/2$. In 
other cases the discussion of this permutation is already implicitly included in our analysis.
In Case I and Case II we also confirm the estimate made for the solar mixing angle $\sin^2 \theta_{12}$ that is bounded from below 
$\sin^2 \theta_{12} \gtrsim 0.276$, while the lower bound in Case III is $1/3$. The upper bound $\sin^2\theta_{12} \lesssim 0.345$ found in
Case IV-P1 and IV-P2 is obeyed as well, see table \ref{table1} and figure \ref{figure1}.

As can be read off from table \ref{table1} for Case II and III, $\Delta \chi^2 = \chi^2_{\footnotesize\mbox{min, IO}}- \chi^2_{\footnotesize\mbox{min, NO}} \approx 3.7$
showing that NO is better compatible with maximal atmospheric mixing $\sin^2\theta_{23}=1/2$. This is simply due to the 
asymmetric $1 \, \sigma$ errors
of the atmospheric mixing angle for NO and IO, see (\ref{nufitresults}).
Concerning the results for Case IV-P2 one might have naively expected that $\chi^2_{\footnotesize\mbox{min, IO}}$ is smaller than  $\chi^2_{\footnotesize\mbox{min, NO}}$, since the value obtained for the atmospheric mixing angle at the best fitting point $\theta_{\footnotesize\mbox{bf}} \approx 0.255$
is larger than $\pi/4$, i.e. $\sin^2\theta_{23} > 1/2$. However, due to the large $1 \, \sigma$ error associated with $\sin^2\theta_{23}$ in the case of NO, for a value
of the latter larger than $(\sin^2\theta_{23})^{\mathrm{bf}}_{\mathrm{NO}}=0.452$, namely $+0.052$ and a smaller one in the case of IO, for values of $\sin^2\theta_{23}$ smaller
than the best fit value $(\sin^2\theta_{23})^{\mathrm{bf}}_{\mathrm{IO}}=0.579$, namely $-0.037$, 
see (\ref{nufitresults}), respectively, 
we find that NO fits slightly better in this case.  

\begin{figure}[t!]
\begin{center}
\begin{tabular}{cc}
\includegraphics[width=0.5\textwidth]{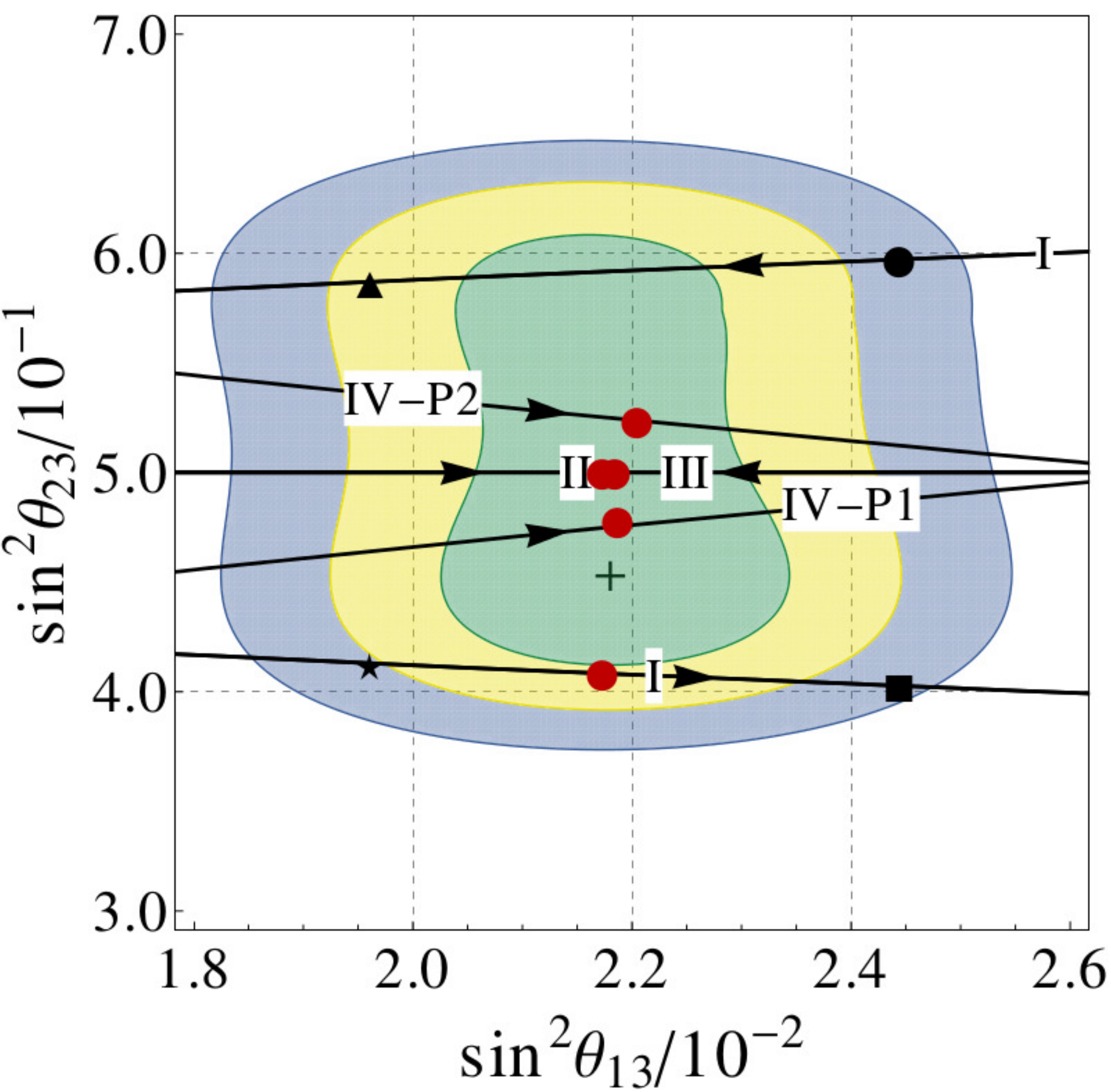} &
\includegraphics[width=0.5\textwidth]{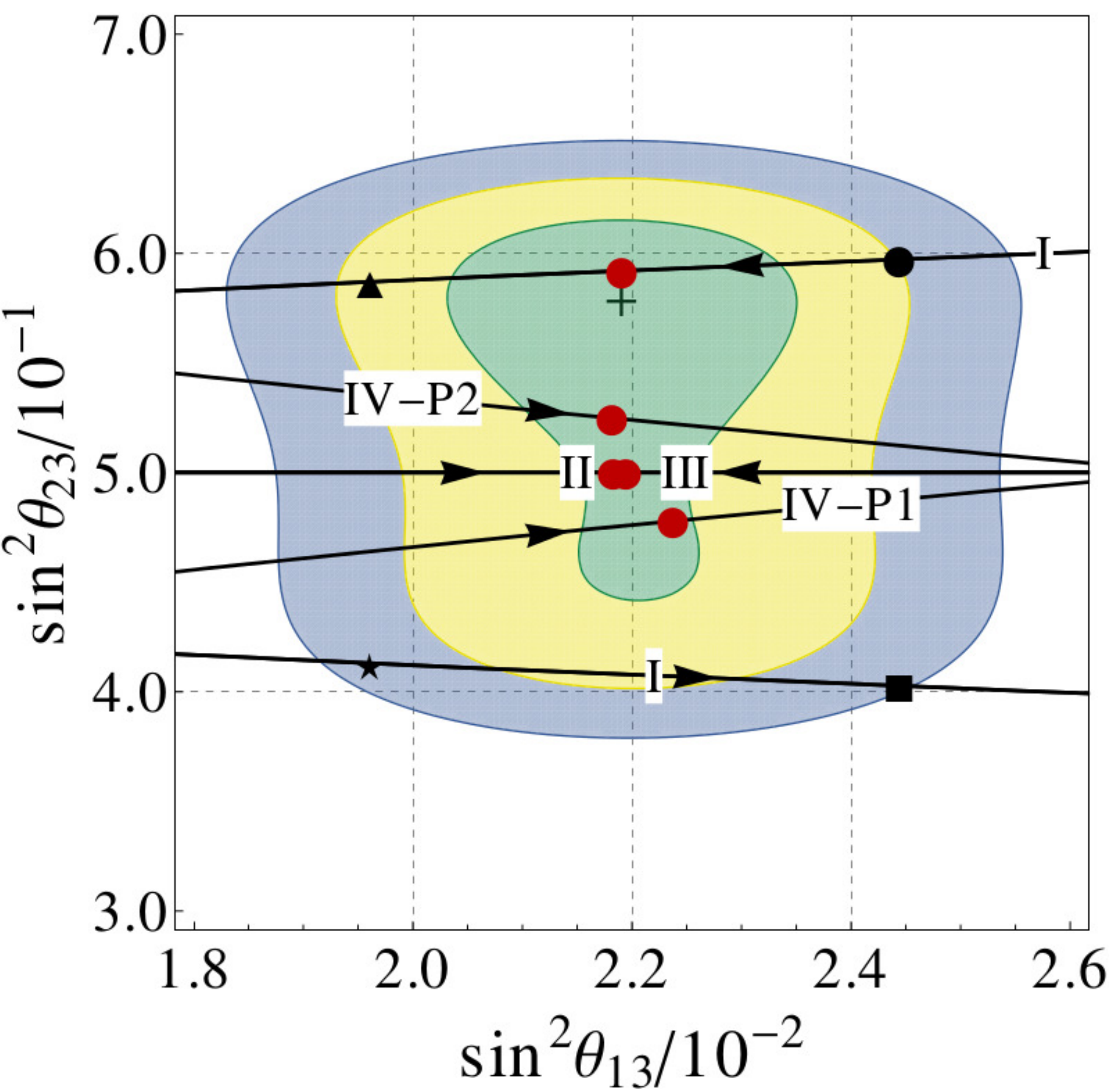}\\
\includegraphics[width=0.5\textwidth]{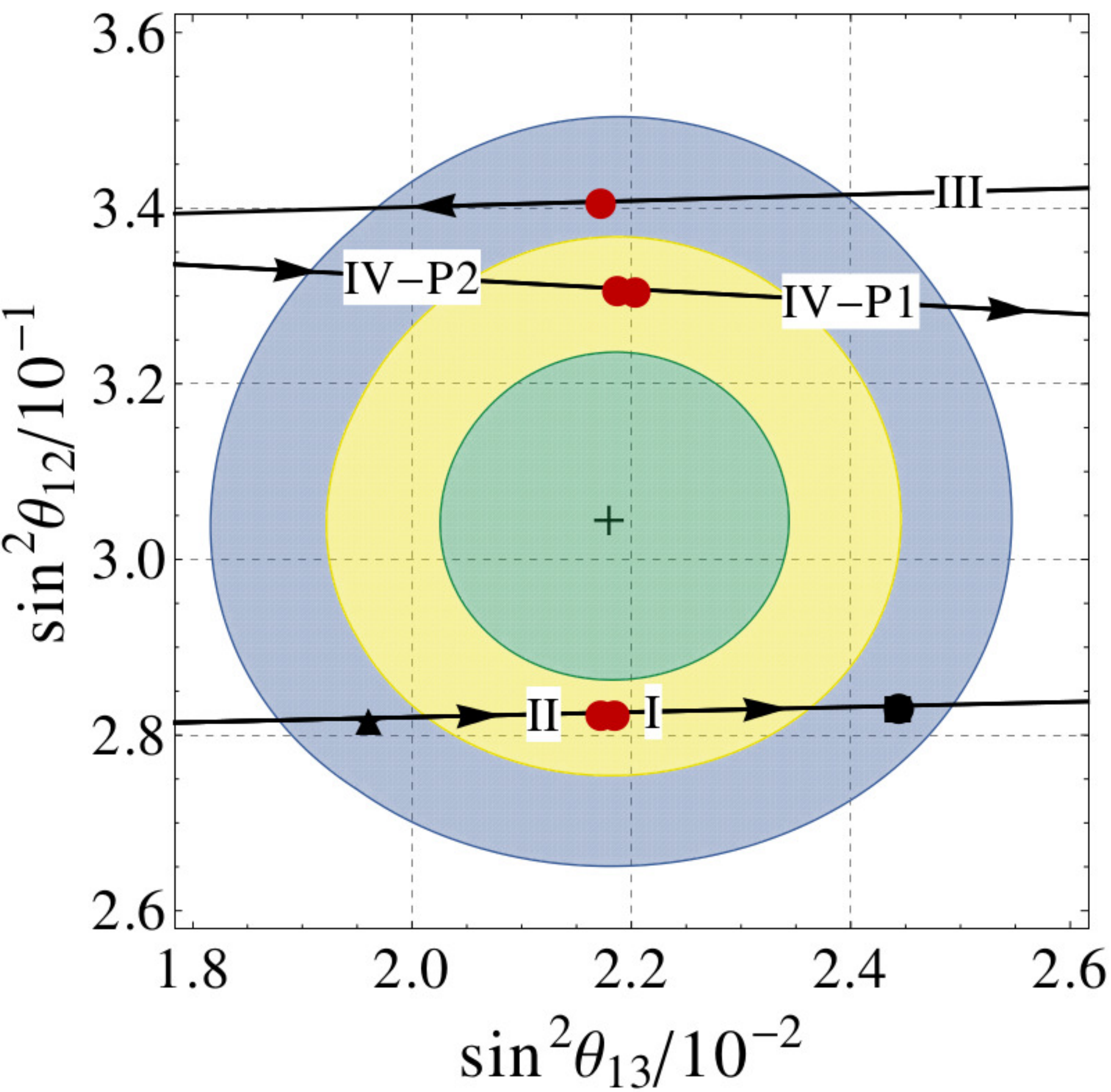} &
\includegraphics[width=0.5\textwidth]{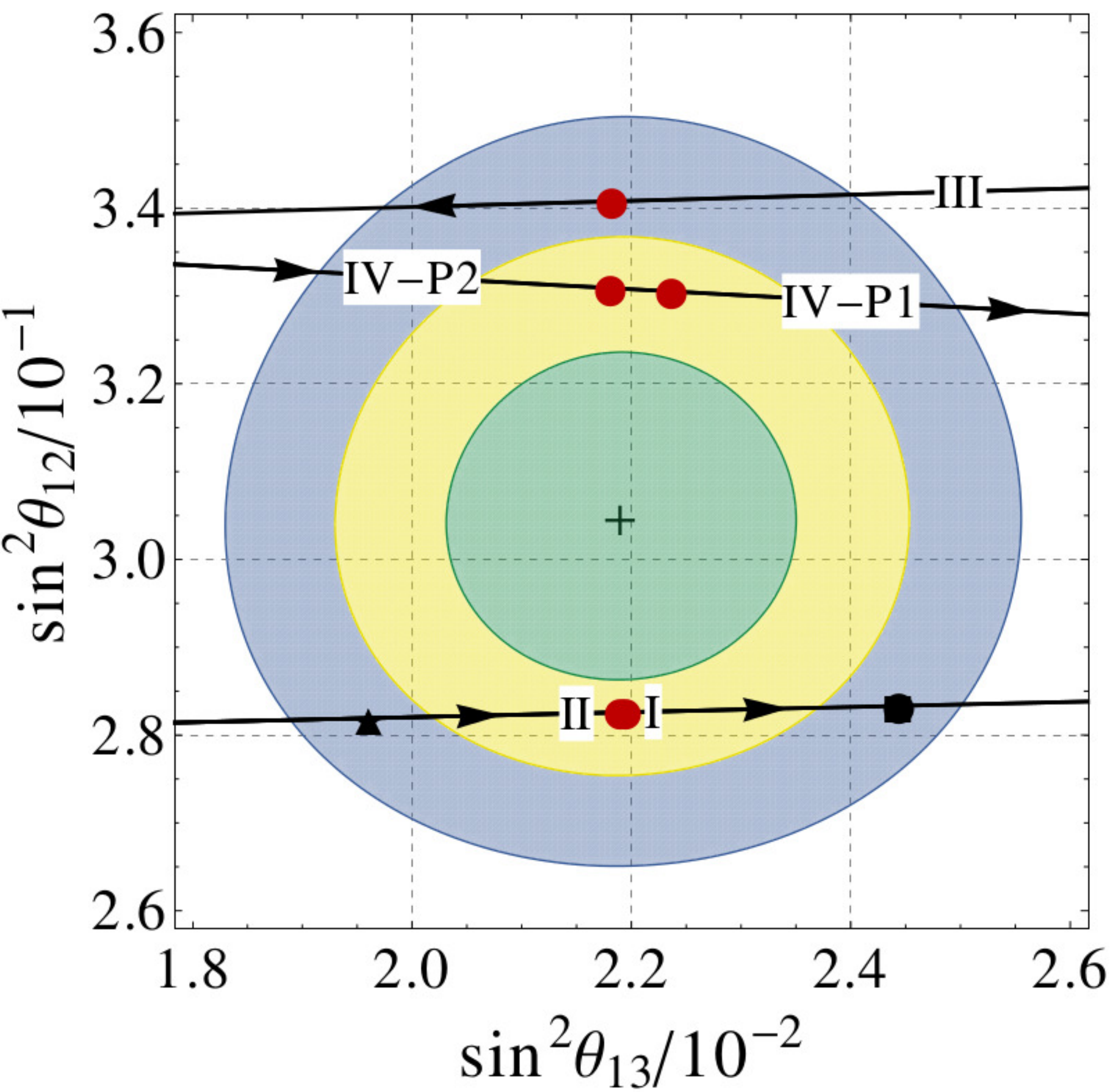}
\end{tabular}
\caption{\small {\it
Results for the mixing angles $\sin^{2}\theta_{23,12}$ with respect to $\sin^2\theta_{13}$ for Case I to III and Case IV-P1 and IV-P2. 
For conventions see caption of figure \ref{figure1}. The two curves for Case I in the upper two plots appear not to be connected
due to the range plotted. Furthermore, note that the curves belonging to Case II and Case III overlap in the $\sin^2\theta_{13}$-$\sin^2\theta_{23}$ plane.
Similarly, the curves for Case I and II as well as for Case IV-P1 and IV-P2, respectively, lie 
on top of each other in the $\sin^2\theta_{13}$-$\sin^2\theta_{12}$ plane.}}
\label{figure2}
\end{center}
\end{figure}

The results found in table \ref{table1} can be nicely visualized in the different $\sin^2\theta_{ij}$ planes, see figures \ref{figure1} and \ref{figure2}.
We plot the experimentally preferred $1 \, \sigma$, $2 \, \sigma$ and $3 \, \sigma$ areas at two degrees of freedom in different colors (green, yellow and blue, respectively)
for a neutrino mass spectrum with NO and IO (left and right panels, respectively) in the $\sin^2\theta_{ij}$ planes using the data sets available in \cite{nufit}.\footnote{
Those for $\sin^2\theta_{13}$-$\sin^2\theta_{12}$ and  $\sin^2\theta_{13}$-$\sin^2\theta_{23}$ are explicitly given in \cite{nufit}, while the one for the third $\sin^2\theta_{ij}$ plane can be simply
constructed by summing up the values of the one-dimensional $\chi^2$ projections for $\sin^2\theta_{12}$ and $\sin^2\theta_{23}$. This is justified, since these two mixing angles are to good approximation uncorrelated. We thank Thomas Schwetz-Mangold for help regarding this point. 
} 
In order not to penalize NO we subtract $\Delta \chi^2=0.97$ in this case so that the minimum value of $\Delta \chi^2$ is zero for both mass orderings.
The experimental best fit values of $\sin^2\theta_{ij}$ are presented by a cross in each plane.
The shaded areas are to be compared with 
the black curves parameterized with $\theta$ shown for Case I through IV-P2. On these curves we indicate the point(s) $\theta_{\footnotesize\mbox{bf}}$ at which the $\chi^2$
function for NO and IO, respectively, is minimized with a red dot. Furthermore, an arrow on each curve marks the direction
of increasing $\theta$ in the interval $0 \leq \theta \leq \theta_{\footnotesize\mbox{bf} (,1)}$. We use the convention that the
arrow always belongs to the closest label I, II, etc. of a certain case. Since the curve representing the results of Case I appears to be disconnected
in the upper plots in figure \ref{figure2} due to the chosen scales of the axes we indicate different values of the parameter $\theta$ with
different symbols on the curve. In figure \ref{figure1} dashed red vertical lines show the lower and upper
bounds on $\sin^2 \theta_{12}$ that exist in the various cases, see (\ref{GeZ5sumrule1}), (\ref{GeZ3sumrule}) and comment below (\ref{GeZ2Z2sumrule1}).
Several curves (partly) overlap in the different planes: in figure \ref{figure1} and the upper two plots in figure \ref{figure2} clearly the curves belonging to Case II and III overlap, 
since both these cases predict maximal atmospheric mixing; furthermore, in the lower plots in figure \ref{figure2} the curves of Case I and II
as well as of Case IV-P1 and Case IV-P2 lie on top of each other, since the relation between the solar and the reactor mixing angles is identical in these
cases. Eventually, we note that in the plots showing $\sin^2 \theta_{23}$ on the vertical axis in figures \ref{figure1} and \ref{figure2}, the fact that the patterns of
Case IV-P1 and IV-P2 are related by the exchange of the second and third rows in the PMNS mixing matrix is clearly visible, since their curves
are symmetric with respect to the value $\sin^2\theta_{23}=1/2$. The curve belonging to Case I itself possesses this property showing that values of 
$\sin^2\theta_{23}$ smaller or larger than 1/2 can be achieved for different choices of the parameter $\theta$. For this reason the best fitting point $\theta_{\mathrm{bf}}$
is at a (very) different position for NO and for IO in figure \ref{figure1} as well as in the upper plots of figure \ref{figure2}. This does not occur in the other
cases.

We note that for Case I there exists a second (local) minimum of the $\chi^2$ function for NO and IO with $\chi^2_{\footnotesize\mbox{min}} \lesssim 27$.
As can be guessed this second minimum for NO is obtained at $\theta_{\footnotesize\mbox{bf}} \approx 2.967$ for $\chi^2_{\footnotesize\mbox{min}} \approx 10.42$
and the mixing angles are very similar to those achieved at the global minimum of IO, $\sin^2 \theta_{12} \approx 0.283$,
$\sin^2 \theta_{13} \approx 0.0217$ and $\sin^2 \theta_{23} \approx 0.592$. In the same vein we find a second (local) minimum of the $\chi^2$ function for IO
at the best fitting point $\theta_{\footnotesize\mbox{bf}} \approx 0.174$ for $\chi^2_{\footnotesize\mbox{min}} \approx 24.52$. The values obtained for the 
mixing angles are practically those of the global minimum of NO, i.e. $\sin^2\theta_{12} \approx 0.283$, $\sin^2\theta_{13}\approx 0.0217$
and $\sin^2\theta_{23}\approx 0.408$, compare table \ref{table1}. In all other cases the global minimum of the $\chi^2$ function that we mention in table \ref{table1}
is the only minimum with $\chi^2_{\footnotesize\mbox{min}}  \lesssim 27$.

A closer look at table \ref{table1} reveals that the solar mixing 
angle differs by up to 20\% between Case I/Case II with
$\sin^2 \theta_{12} \approx 0.283$ and Case III/Case IV-P1 and IV-P2 where larger
values of $\sin^2\theta_{12}$ are obtained. The experiment JUNO \cite{refjuno} will be able
to reduce the error on the best fit value of $\sin^2\theta_{12}$ to $\sim$ 0.7\% at the $2 \, \sigma$ level, thus allowing for a discrimination 
among  Case I/Case II and Case III/Case IV-P1 and IV-P2. According to the 
RENO-50 collaboration their planned experiment can achieve a similar reduction
of the error \cite{refRENO50}.
For the atmospheric mixing angle, no distinction is possible
among Case II and III, since $\theta_{23}$ is maximal in both cases. 
However, the predictions for Case I and Case IV-P1/Case IV-P2 considerably differ:
for NO (IO) the atmospheric mixing angle is smaller (larger) in Case I than
in Case IV-P1 (Case IV-P2). This difference is
large enough to be possibly distinguished in the experiment NO$\nu$A \cite{Ayres:2004js}.
This experiment can also help in measuring the Dirac phase $\delta$, especially, if running
in both, the neutrino and the anti-neutrino, modes so that a discrimination between Case II/Case III where $\delta$
is maximal and Case I/Case IV-P1 and IV-P2 with $\delta=0, \, \pi$ might be possible. 
In contrast, the predictions for $\theta_{13}$ are almost the same in Case I through Case IV-P1 and IV-P2
 and, hence, it is unlikely that they can be distinguished at future neutrino facilities.

Furthermore, we mention the outcome of our $\chi^2$ analysis for the additional case for $G_e=Z_5$ 
where we can choose as representative generators of the residual symmetries $(Q,Z,X)=(T, S, X_0)$.
We remind the reader that this case fails in giving a good fit to the solar mixing angle.
We find two best fitting points $\theta_{\footnotesize\mbox{bf}, 1} \approx 0.283 (4)$
and  $\theta_{\footnotesize\mbox{bf}, 2} \approx 2.858$ for NO (IO) whose sum approximately equals $\pi$. Like in the cases above, this is 
expected from some symmetry of the formulae for the solar and the reactor mixing angles. The minimal $\chi^2$
values for NO and IO are at these two points $\chi^2_{\footnotesize\mbox{min, NO}} \approx 14.08$ and $\chi^2_{\footnotesize\mbox{min, IO}} \approx 17.83$,
respectively. Thus, also here $\Delta \chi^2$, the difference of $\chi^2_{\footnotesize\mbox{min, IO}}$ and $\chi^2_{\footnotesize\mbox{min, NO}}$, is about 3.7, 
since the atmospheric mixing angle is fixed to be maximal by this pattern. The values obtained for the other mixing angles are $\sin^2 \theta_{12} \approx 0.260$
 and $\sin^2\theta_{13} \approx 0.0216 (7)$ and the Dirac phase $\delta$ is maximal. In particular, we find for 
 $\theta=\theta_{\footnotesize\mbox{bf}, 1}$, $\sin\delta=-1$ and $J_{CP} \approx - 0.0315 (6)$ for NO (IO) and for the other value $\theta=\theta_{\footnotesize\mbox{bf}, 2}$, $\sin\delta=1$
and a positive sign also for $J_{CP}$. Like in the cases presented in table \ref{table1} the two Majorana phases $\alpha$ and $\beta$ are trivial.

\vspace{0.1in}
All patterns that can accommodate the experimental data well for a certain value of $\theta$ predict trivial Majorana phases independently of $\theta$.
However, this is not a general result of the scenario with the flavor group $A_5$ and a CP symmetry. Instead  
there are also patterns that lead to three non-trivial CP phases which depend on the parameter $\theta$ and that are also non-trivial 
at the best fitting point(s) $\theta_{\footnotesize\mbox{bf}}$. For $G_e$ being a Klein group, there exists a further category which can
accommodate three non-trivial mixing angles, see discussion at the beginning of subsection \ref{sec33}. This category also predicts non-trivial and non-maximal Dirac and Majorana phases. 
A representative is $(\left\{Q_1, Q_2\right\},Z,X)=(\left\{ S , T^2 S T^3 S T^2 \right\}, S T^3 S T^2 S, T^4 S T X_0)$.
The third column of the PMNS mixing matrix has then components that are the same as in 
(\ref{GeZ2Z2vector}) up to permutations. Thus, the reactor and the atmospheric mixing angles do not depend 
on $\theta$ and are outside the experimentally preferred $3 \, \sigma$ ranges. As a consequence, 
the $\chi^2$ values for NO and IO are (much) larger than 100 and it appears in general difficult to achieve good agreement with the 
data, even if corrections to the leading order results presented here are expected to exist in an explicit model.

\section{Summary}
\label{summ}

We have discussed a scenario for three Majorana neutrinos with the flavor group $A_5$ and a CP symmetry. 
These symmetries are broken to residual groups $G_e= Z_3$, $Z_5$ or $Z_2 \times Z_2$ and to $G_\nu = Z_2 \times CP$
in the charged lepton and neutrino sectors, respectively. As a consequence, lepton mixing angles as well as CP phases can be 
predicted in terms of a single free parameter $\theta$ that varies between $0$ and $\pi$. We have comprehensively 
studied all possible residual groups $G_e$, possible choices of $Z_2$ being a subgroup of $A_5$ and CP symmetries
that can be consistently combined. We have shown explicitly that the automorphisms corresponding to the CP transformations 
we consider are class-inverting and involutive and, furthermore, that these are the only automorphisms of $A_5$ with such properties. 
Performing a detailed analytical and numerical study, we found that only four mixing patterns exist (two for $G_e=Z_5$ and 
one each for the other two choices, $G_e=Z_3$ and $G_e=Z_2 \times Z_2$) that can accommodate the experimental data
on lepton mixing angles at the $3 \,\sigma$ level or better for a particular value of the parameter $\theta$. The reactor mixing 
angle is usually accommodated very well, while the solar one is bounded from below by three of the four patterns
($\sin^2 \theta_{12} \gtrsim 0.276$ for Case I and Case II and $\sin^2 \theta_{12} \gtrsim 1/3$ for Case III, since this pattern has a TM column) 
and bounded from above, $\sin^2 \theta_{12} \lesssim 0.345$, by the pattern called Case IV-P1/Case IV-P2. Interestingly enough,
two of the four patterns, Case II and Case III, predict a maximal Dirac phase $\delta$ (together with maximal atmospheric mixing),
while the other two patterns lead to $\sin\delta=0$ (and $\theta_{23}$ is in general non-maximal). Common to all patterns is the prediction of trivial Majorana phases. 
Thus, in two of the four setups an accidental CP symmetry, common to the charged lepton and neutrino
sectors, is present. Furthermore, we mention the existence of one further case for the choice $G_e=Z_5$ that can be obtained by using as generators 
of the residual symmetries $(Q,Z,X)=(T,S,X_0)$. This case fits the experimental data well to a certain extent. It mainly fails because of a too small value 
of the solar mixing angle, $\sin^2\theta_{12} \approx 0.260$.
However, this could be easily reconciled with the experimental results in an explicit model with small corrections. The further predictions of this pattern are practically
identical to those of Case II, i.e. the Dirac phase and the atmospheric mixing angle are maximal, the two Majorana phases are trivial and the reactor mixing angle
can be accommodated well. As mentioned, all patterns that are preferred by experimental data turn out to lead to trivial Majorana phases. However, this is not a generic
feature of our scenario with the flavor group $A_5$ and a CP symmetry, but occurs, because none of the patterns with non-trivial Majorana phases admits a
reasonable fit to the data on lepton mixing angles. Interestingly enough, the absence of non-trivial Majorana phases and the correlation
between having a maximal Dirac phase and maximal atmospheric mixing or a trivial Dirac phase and non-maximal $\theta_{23}$ for patterns that fit the
data well have already been observed in analyses of scenarios with the flavor group  $G_f=S_4$ (or $G_f=A_4$) and a CP symmetry \cite{GfCPgeneral,GfCPmodels}.

\vspace{0.1in}
{\bf Note added}: At the very final stages of the completion of this work a paper \cite{Li:2015jxa} dealing with the
same topic appeared on the arXiv. Our results agree with those obtained in \cite{Li:2015jxa}. However,
the $\chi^2$ analysis in this work has been performed using results of a different global fit \cite{nufit}. 
In addition, the authors of \cite{Li:2015jxa} display results for neutrinoless double beta decay and some model realization.

\section*{Acknowledgements}

We thank Thomas Schwetz-Mangold for discussions on the results of \cite{nufit} and the data sets available at the corresponding webpage. A.Di I. and D.M. acknowledge MIUR for financial support under the program Futuro
in Ricerca 2010 (RBFR10O36O). C.H. is grateful to the organizers of the workshop ``XVI International Workshop on Neutrino Telescopes" (02.03.-06.03.2015, Venice, Italy) where parts of this work have been presented.

\appendix

\mathversion{bold}
\section{Comments on automorphisms of $A_5$}
\mathversion{normal}
\label{appA}

The CP transformations in (\ref{formX}) all correspond to inner automorphisms, see (\ref{autoX}). 
We can check that these are class-inverting \cite{ChenCP} and involutive. The latter feature is required, since $X$ has to be a symmetric matrix in the flavor space, as explained below (\ref{Xcon}). Both properties are in particular given, if the twisted Frobenius-Schur indicator $\epsilon_{\iota} (\mathrm{{\bf r}})$ of an automorphism $\iota$ is $+1$ in all
irreps. {\bf r}. The definition of
 $\epsilon_{\iota} (\mathrm{{\bf r}})$ in the case at hand is
\begin{equation}
\label{twistedFSindicator}
\epsilon_{\iota} ({\mathrm{{\bf r}}}) = \frac{1}{60} \, \sum_{g \in A_5} \chi_{\mathrm{{\bf r}}} (g \, {}^{\iota}g)
\end{equation}
with $\chi_{\mathrm{{\bf r}}} (h)$ being the character of the element $h$ and ${}^{\iota}g$ the image of the element $g$ under the automorphism $\iota$.
 An easy way to check \cite{FSind_comp} that $\epsilon_{\iota} (\mathrm{{\bf r}})=+1$ for all ${\bf r}$  is to verify the equality
\begin{equation}
\label{reltwistedFSone}
\sum_{\mathrm{{\bf r}}} \chi_{\mathrm{{\bf r}}} (e) =  \left| \left\{ g \in A_5 \mid {}^{\iota}g=g^{-1} \right\} \right|
\end{equation}
with $\chi_{\mathrm{{\bf r}}} (e)$ being the character of the neutral element $e$ in the irrep {\bf r}, i.e. we sum over the dimensions of all
irreps. of $A_5$ on the left-hand side of (\ref{reltwistedFSone}). 
For the trivial automorphism, see (\ref{autoX0}), the equality in (\ref{reltwistedFSone}) is obvious:
 the left-hand side is the sum of the dimensions of the irreps. {\bf r} that is 16, while the right-hand side is the number of elements of the group $A_5$ that fulfill $g=g^{-1}$. Clearly
these are all $Z_2$ generating elements together with the neutral element. Thus, these are 16 elements. 
For the automorphisms corresponding to choices of $X$ other than $X_0$, see (\ref{formX}), we can do this computation using GAP \cite{gap} and also verify in these
cases that the twisted Frobenius-Schur indicator for all irreps {\bf r} is $+1$. Thus, all CP transformations in (\ref{formX}) correspond to 
class-inverting, involutive automorphisms. 
Furthermore, we can compute the twisted Frobenius-Schur indicator for all
other automorphisms $\iota$ and in all representations {\bf r} of $A_5$ and find that none of the other ones fulfills $\epsilon_{\iota} ({\mathrm{{\bf r}}}) =\pm 1$
for all {\bf r}. So, none of the other automorphisms of $A_5$ is class-inverting and involutive.

\mathversion{bold}
\section{Convention for lepton mixing parameters and CP invariants}
\mathversion{normal}
\label{appB}

We use the following convention for the PMNS mixing matrix
\begin{equation}
U_{PMNS} = \tilde{U} \, \mbox{diag} \left( 1, e^{i \, \alpha/2}, e^{i \, (\beta/2 + \delta)} \right)
\end{equation}
with $\tilde{U}$ being defined, similar to the Cabibbo-Kobayashi-Maskawa mixing matrix \cite{pdg}, 
\begin{equation}
\!\!\!\!\! \tilde{U} = \left(\begin{array}{ccc}
        1 & 0 & 0 \\
        0 & \cos\theta_{23} & \sin\theta_{23} \\
        0 & -\sin\theta_{23} & \cos\theta_{23} 
 \end{array} \right)\,
\left(\begin{array}{ccc}
        \cos\theta_{13} & 0 & \sin\theta_{13} \, e^{-i \, \delta} \\
        0 & 1 & 0 \\
        -\sin\theta_{13} \, e^{i \, \delta} & 0 & \cos\theta_{13} 
\end{array} \right) \,
\left(\begin{array}{ccc}
        \cos\theta_{12} & \sin\theta_{12} & 0 \\
        -\sin\theta_{12} & \cos\theta_{12} & 0 \\
        0 &  0 & 1 
 \end{array} \right) \; .
\end{equation} 
The mixing angles $\theta_{ij}$ are taken to be in the interval between $0$ and $\pi/2$.
The Dirac phase $\delta$ as well as the two Majorana phases $\alpha$ and $\beta$ can assume values between $0$ and $2 \ \pi$. The Dirac phase $\delta$
can be extracted using the Jarlskog invariant $J_{CP}$ \cite{JCP}
\begin{eqnarray}\nonumber
 J_{CP} &=&  \mathrm{Im} \left( U_{PMNS, 11} U^\star_{PMNS, 13}U^\star_{PMNS, 31}U_{PMNS, 33} \right)
 \\
  &=& \frac 18 \, \sin 2\theta_{12} \sin 2\theta_{23} \sin 2\theta_{13}\cos\theta_{13} \sin\delta \; .
\end{eqnarray}
Similar invariants, called $I_1$ and $I_2$, can be defined for the Majorana phases
\begin{eqnarray}
 I_1 &=& \mathrm{Im} \left( U_{PMNS, 12}^2 (U^\star_{PMNS, 11})^2 \right) = \sin^2\theta_{12}\cos^2\theta_{12}\cos^4\theta_{13}\sin\alpha \; ,
\\ 
 I_2 &=& \mathrm{Im} \left( U_{PMNS, 13}^2  (U^\star_{PMNS, 11})^2 \right) = \sin^2\theta_{13}\cos^2\theta_{12}\cos^2\theta_{13}\sin\beta \; .
\end{eqnarray}
Notice that the Dirac phase $\delta$ has a physical meaning only if all mixing angles are different from $0$
and $\pi/2$. Analogously, the vanishing of the invariants $I_{1,2}$ only implies
$\sin\alpha = 0$, $\sin\beta = 0$, if solutions with $\sin 2\theta_{12} = 0$, $\cos \theta_{13} = 0$ or $\sin 2\theta_{13} = 0$, $\cos \theta_{12} = 0$ are
discarded. Furthermore, notice that one of the Majorana phases becomes unphysical, if the
lightest neutrino mass vanishes.

 

\end{document}